\documentclass[10pt]{bmc_article}    

\usepackage{graphicx}
\usepackage{cite} 
\usepackage{url}  
\usepackage{ifthen}  
\usepackage{multicol}   
\usepackage[utf8]{inputenc} 
\usepackage{rotating}
\usepackage{epsfig}
\usepackage{amsmath}
\usepackage{array}
\newboolean{publ}

\newenvironment{bmcformat}{\baselineskip20pt\sloppy\setboolean{publ}{false}}{\baselineskip20pt\sloppy}

\def\be{\begin{equation}} 
\def\ee{\end{equation}} 
\newcommand \bea {\begin{eqnarray} } 
\newcommand \eea {\end{eqnarray}} 
\newcommand{\nn} {\nonumber}

\begin{document}
\begin{bmcformat}

\title{Gene autoregulation via intronic microRNAs and its functions}

\author{Carla Bosia\correspondingauthor$^{1,2,\dagger}$ %
         \email{Carla Bosia\correspondingauthor - carla.bosia@hugef-torino.org}%
       \and
         Matteo Osella\correspondingauthor$^{3,4,\dagger }$ %
         \email{Matteo Osella\correspondingauthor - matteo.osella@upmc.fr}%
	\and
	  Mariama El Baroudi$^{5}$ %
	\email{Mariama El Baroudi - mariama.elbaroudi@iit.cnr.it}%
	 \and
	  Davide Cor\'{a}$^{2,6}$ %
	\email{Davide Cor\'{a} - davide.cora@ircc.it}%
and 
 	Michele Caselle$^{2,7}$ %
	\email{Michele Caselle - caselle@to.infn.it}%
}

\address{%
     \iid(1) Human Genetic Foundation, Molecular Biotechnology Center, University of Torino, V. Nizza 52 a, I-10126, Torino, Italy.  
     \iid(2)  Center for Complex Systems in Molecular Biology and Medicine, University of Torino, V. Accademia Albertina 13, I-10100 Torino, Italy.    
	\iid(3) Genomic Physics Group, FRE 3214 CNRS “Microorganism Genomics”, France.
\iid(4) Universit\'{e} Pierre et Marie Curie, 15 rue de L'\'{E}cole de M\'{e}decine, Paris, France.
\iid(5) LISM (Laboratory for Integrative System Medicine), CNR,  Via G. Moruzzi 1, Pisa, Italy.
\iid(6) Systems Biology Lab, Institute for Cancer Research and Treatment (IRCC), School of Medicine, University of Torino, Str. Prov. 142, Km. 3.95, Candiolo I-10060 Torino, Italy.
 \iid(7) Dipartimento di Fisica Teorica and INFN, University of Torino, V. Pietro Giuria 1, I-10125 Torino Italy.\\
$\dagger$ Equal contributors
}%

\maketitle


\begin{abstract}

{\bf BACKGROUND}
MicroRNAs, post-transcriptional repressors of gene expression, play a pivotal role in gene regulatory networks. They are involved in core cellular 
processes and their dysregulation is associated to a broad range of human diseases. 
This paper focus on a minimal microRNA-mediated regulatory circuit, in which a protein-coding gene (host gene) is targeted by a microRNA located inside one of its introns. 

{\bf RESULTS}
Autoregulation via intronic microRNAs is widespread in the human regulatory network, as confirmed by our bioinformatic analysis, and can perform several regulatory tasks despite its simple topology. 
Our analysis, based on analytical calculations and simulations, indicates that this circuitry alters the dynamics of the host gene expression, 
can induce complex responses implementing adaptation and Weber's law, and efficiently filters fluctuations propagating from the upstream network to the host gene. A fine-tuning of the circuit parameters can optimize each of these functions. Interestingly, they are all related to gene expression homeostasis, in agreement with the increasing evidence suggesting a role of microRNA regulation in conferring robustness to biological processes. In addition to model analysis, we present a list of bioinformatically predicted candidate circuits in human for future experimental tests.  

{\bf CONCLUSIONS}
The results presented here suggest a potentially relevant functional role for negative self-regulation via intronic microRNAs, in particular as a homeostatic control mechanism of gene expression. 
Moreover, the map of circuit functions in terms of experimentally measurable parameters, resulting from our analysis, can be a useful guideline 
for possible applications in synthetic biology. 
      
\end{abstract}

\ifthenelse{\boolean{publ}}{\begin{multicols}{2}}{}


\section*{Background}

microRNAs (miRNAs) are small (about $22$ nucleotides) single-strand RNAs able to interfere post-transcriptionally with the protein production of their targets. 
Targeting a vast proportion of protein-coding genes~\cite{Lewis2005,Flynt2008,Friedman2009}, miRNA-mediated regulation composes an important layer in gene regulatory networks. 
The implication of miRNAs in several core cellular processes~\cite{Ambros2004,Bartel2004,Bushati2007,Stefani2008} as well as in many human diseases~\cite{Alvarez-Garcia2005,Esquela-Kerscher2006} 
further confirms their biological importance.\newline
Approximately half of the miRNA genes can be found in intergenic regions (between genes), whereas the intragenic miRNAs (inside genes) are predominantly 
located inside introns and usually oriented on the same DNA strand of the host gene~\cite{Hinske2010} (a trend further confirmed by our bioinformatic analysis shown in a following section). 
Intergenic miRNA genes present their own  promoter region~\cite{Li2007,Saini2007} and their expression is expected to be regulated by the same molecular mechanisms that control the 
expression of protein-coding genes. On the other hand, 
experimental and computational results are consistent with the idea that same-strand intronic miRNAs are co-transcribed with their 
host gene~\cite{Rodriguez2004,Baskerville2005,Lin2006,Kim2007,Ma2011}, and then processed to finally become mature functional miRNAs~\cite{Kim2009,Morlando2008} 
(although exceptions to this common scheme of co-transcripton  have been reported~\cite{Ozsolak2008,Monteys2010,He2012}). \newline
The host-miRNA co-expression can have a specific functional role. In fact, an intronic miRNA can support the function of its host gene by silencing genes that are functionally antagonistic 
to the host \cite{Barik2008}, or more generally act synergistically with the host by coordinating the expression of genes with related functions \cite{Lutter2010}.\newline
In addition to this ``cooperative'' miRNA-host relation, different studies showed that intronic miRNAs can directly regulate the expression of their host gene, 
establishing a negative feedback regulation~\cite{Hinske2010,Tsang2007,Megraw2009}. In particular, instances of negative autoregulatory feedbacks via intronic miRNAs were firstly 
found by expression analysis in human~\cite{Tsang2007}. 
More recently, two independent large-scale bioinformatic analysis, based on different algorithms of target prediction, claimed    
that the occurency of intronic miRNA-mediated self-loops (iMSLs) in the human regulatory network is significantly higher than expected by 
chance~\cite{Hinske2010,Megraw2009}. The over-representation of such regulatory module can be interpreted as a sign of evolutionary positive 
selection that has led to an accumulation of a specific topology able to perform useful elementary regulatory tasks~\cite{Alon2007}. 
In addition, two iMSL circuits have been confirmed experimentally: regulation of EGFL7 by its intronic miRNA miR-126~\cite{Sun2010,Nikolic2010} and 
regulation of ARPP-21 by miR-128b~\cite{Megraw2009}. Both regulations were associated to relevant biological functions, the former playing a role in 
cancer proliferation~\cite{Sun2010}, while the latter in vertebrate brain physiology~\cite{Megraw2009}.\newline
The combination of all these pieces of evidence suggests that iMSLs are an often exploited and presumably functionally relevant regulatory circuitry. 
The open question concerns the peculiar functions that an iMSL can accomplish and that could have thus driven their 
pervasive spreading in the human regulatory network. Moreover, it would be interesting to understand what specificities of post-transcriptional 
autoregulation by miRNAs can make them better suited to fullfil certain tasks with respect to the trascriptional self-regulation, so widely used in 
bacteria~\cite{Alon2006}.
In this paper we address these questions by modeling the dynamical and stochastic behaviour of the iMSL circuit and comparing its properties to those of 
alternative regulatory strategies such as constitutive expression and transcriptional self repression.\newline
Our results show that, despite of its minimal topology, the iMSL circuitry can implement different biological functions. It can speed-up the host gene protein 
production in response to an activating signal, while delaying its switching-off kinetics when the activation drops; it can buffer fluctuations coming 
from the upstream network, and generate complex behaviours like a host gene expression response obeying ``Weber's law" (i.e. the magnitude of the response 
depends only on the fold change of the input signal). 
While these different functions can be optimized individually, by tuning parameters 
like molecular production/degradation rates, it will be shown that they all represent different ways of making the host gene expression robust to 
external fluctuations. Therefore, autoregulation via intronic miRNAs can generally represent an efficient homeostatic control  of the host gene expression, 
in agreement with the observation that miRNAs are often involved in signaling networks to ensure homeostasis and gene expression robustness~\cite{Inui2010,Li2009,Li2006,Ebert2012}. 
In addition to model analysis, we present our own bioinformatical search for iMSLs in human to further assess 
their statistical over-representation and to propose the best predicted candidates to eventual future experimental tests.\newline
Besides the understanding of the role of endogenous iMSLs, our results can be useful for the growing field of synthetic biology~\cite{Haynes2009,Khalil2010}, 
which has succesfully started to make use of RNA-based post-transcriptional regulations~\cite{Deans2007,Rinaudo2007}. 
The function-topology map presented in this paper can contribute to draw up the manual of biological circuits 
that carry out specific functions for synthetic engineering, adding a simple and efficient wiring strategy that can increase systems' robustness in different conditions. 
A synthetic realization of an iMSL has been indeed recently produced and proven to be effective in reducing the expression dependency on gene dosage~\cite{Bleris2011}. 
Therefore, the potential additional functions we will show associated to iMSLs could be tested in the near future.

\section*{Results and discussion}

\subsection*{Outline of the model.}

We are interested in a model of iMSLs that can capture the fundamental properties of the circuit, but simplified enough 
to avoid the introduction of too many free parameters that would make an exploration of the parameter space unfeasible.
In this view, the essential steps of transcription, translation, degradation and interactions between genes are taken 
into account as summarized in Figure 1A. 
The host gene is assumed to be under the control of an activating transcription factor (TF) with concentration $q$,  
in order to study the dynamical and stochastic properties of the circuit in presence of upstream input signals. 
The activation is modeled, as usual in this type of descriptions~\cite{Alon2006,Bintu2005}, representing the transcription 
rate of the target as a Michaelis-Menten function of TF concentration:

\begin{equation}
 k_r(q) =\frac {k_r q} {h_{r} +q}.
\label{activhill}
\end{equation}

However, the analysis can be straightforwardly extended to the case of a Hill function (substituting $q$ with $q^n$ and $h_r$ with $h_{r}^{n}$), if in presence of cooperativity. \newline
On the other hand, there is currently no standard and clearly tested strategy for modeling miRNA-mediated repression. First of all, miRNAs can exert their action repressing translation 
or inducing degradation of their target mRNAs~\cite{Valencia-Sanchez2006}. 
We construct our model supposing an action on target translation. While most of the results shown in this paper are independent of this choice, some dynamical properties of the circuit can actually change 
if miRNA action is mainly due to induction of mRNA degradation. This issue is discussed in more detail in Additional file 1.\newline 
A phenomenological description based on nonlinear functions has been proven to be effective in modeling RNA interference in mammals~\cite{Cuccato2011}, 
and was previously applied in computational analysis~\cite{Komorowski2009,Osella2011}. Along these lines, 
we assume that miRNA regulation makes the target translation rate $k_p(s)$ a repressive Michaelis-Menten-like function of the number of miRNAs ($s$):

\begin{equation} 
k_p(s) =\frac {k_p} {1 + \frac{s}{h}}.
\label{reprhill}
\end{equation}

With the regulatory interactions defined in Equations \ref{activhill} and \ref{reprhill}, it is possible to represent the dynamics of the circuit in Figure 1A by a set of differential equations:

\begin{eqnarray}
\frac{d {r}}{{d t}} & =& k_{r}(q)  - g_{r} r \nonumber \\
\frac{d {s}}{{d t}} & =& k_{r}(q)  - g_{s} s \nonumber \\
\frac{d {p}}{{d t}} & =& k_{p}(s)~ r - g_{p} p,
\label{det-model}
\end{eqnarray}

where $r$ and $p$ are the levels of host gene mRNA and protein products, while $s$ is the level of miRNAs. 
As discussed in the introduction, intronic miRNAs (same-strand with the host) are expected to be co-transcribed with their host gene,  
therefore their production rate $k_{r}(q)$ has the same dependence on the input TF level. \newline
A different representation was introduced in the context of bacterial sRNA regulation~\cite{Levine2007,Shimoni2007,Mehta2008} and subsequently 
applied  with slight modifications to eukaryotic miRNA regulation~\cite{Mukherji2011}. In this representation, the degree of catalicity, 
i.e. the ability of miRNA to affect multiple mRNAs without being degraded, was parametrized explicitly~\cite{Levine2007}. 
The use of an effective phenomenological function (like the one in Equation \ref{reprhill}) implicitly assumes a catalytic action, as commonly believed for miRNAs~\cite{Cuccato2011}. 
The relations between different possible models are discussed more precisely in the supporting information (Additional file 1), where it is shown that most of the results that will be presented in the following are essentially independent on the modeling strategy, provided that certain generic conditions on the parameters are satisfied.\newline
In an analogous manner, it is possible to model the two circuits that we will use for comparison: a gene simply activated by the TF (sTF) 
without any feedback regulation (scheme in Figure 1B) and a transcriptional self-loop (tSL), in which the negative feedback is realized through transcriptional repression 
(scheme in Figure 1C). The properties of each circuit will be compared using a so called \textit{mathematically controlled comparison}~\cite{Alon2006}:  all the common parameters will be kept to equivalent values, 
constraining the remainders so as to achieve the same steady state of protein concentration.\newline  
A deterministic description based on ordinary differential equations can effectively describe the mean kinetic behaviour of genetic circuits, thus its predictions 
can be tested with experiments based on averages over cell populations. In fact, equivalent mathematical treatments have correctly predicted the dynamic features of 
several endogenous and synthetics circuits~\cite{Alon2006,Alon2007}. 
However, since gene expression is inherently a stochastic process~\cite{Kaern2005,Maheshri2007,Raj2008}, we will also make use of a stochastic description based on a master 
equation approach, that has Equations~\ref{det-model}  as a ``mean-field'' limit (complete model in Additional file 1). 
To compare the stochastic properties and the noise susceptibility of the three regulatory  
strategies in Figure 1, we calculated analytically the relative fluctuations in protein level $p$ at steady state and confirmed our results with Gillespie simulations 
(see the Methods section for details on simulations).

\subsection*{Response times to external signals are altered by autoregulation via intronic microRNAs}

The response of a transcriptional unit to a stimulus, such as a change in a TF concentration,  is steered by the lifetime of its mRNA and protein products. 
A fast protein turnover speeds up the kinetics, but with a consequent high metabolic cost, while in the case of long-living proteins the timescale of changes in concentration can be comparable to the cell cycle time~\cite{Alon2006,Rosenfeld2002}, which can be of several hours. 
However, the dynamics of a gene expression also depends strongly on the regulatory circuitry in which the gene is embedded. 
For example, it has been proven that negative transcriptional self regulation (like the one in Figure 1C) and incoherent feed-forward loops  speed up 
the expression rise-time after induction~\cite{Rosenfeld2002,Mangan2006}, while coherent feed-forwad loops introduce delays~\cite{Mangan2003}.\newline
We address in this section the question of how the host gene kinetics is changed by being a target of its intronic miRNA. 
To this aim, we consider two opposite simplified situations: (i) a sudden activating signal that fully saturates the promoter, and (ii) the opposite case of an istantaneous drop of the activating signal that completely switches off transcription. 
Case (i) can be studied assuming that at $t=0$ the transcription rate $k_{r}(q)$ switches from its maximum value $k_r$ to zero, and measuring the 
response time $T_{ON}$ defined as the time needed to reach half of the final protein steady-state. 
In other words, we integrate numerically Equations~\ref{det-model} to calculate the time $T_{ON}$ such that $p(T_{ON})/p_{ss} =0.5$ (where $p_{ss}$ is the final steady-state protein level), 
starting from the condition $r(0)=s(0)=p(0)=0$. 
In case (ii), in which we assume a drop of the activating signal at $t=0$, we can similarly define a 
response time $T_{OFF}$ looking at the decrease of  $p(t)$ after a switch of the transcription rate from $k _{r}$ to zero at time $t=0$.
The same analysis is performed on a sTF (scheme in Figure 1B) and a tSL (scheme in Figure 1C) for comparison. 
The response time $T_{0}$ of the simple transcription unit sTF is used as a normalization, 
since  $T_{ON(OFF)}/T_{0}$ is a measure of how much a circuit can alter the response time with respect to an unregulated gene.\newline
Many previous analyses of genetic circuit dynamics have assumed short-living mRNAs with respect to proteins~\cite{Rosenfeld2002,Mangan2003,Mangan2006}. Within this assumption, 
the mRNA dynamics can be neglected 
since the timescales are governed by the protein kinetics. While this is usually a safe approximation in bacteria, in eukaryotes the phenomenology can be more complex.
In mammals, the mRNA half-life can range from minutes to about 24 hours~\cite{Fan2002,Chen2008}, with  
typical values in the range of $5-10$ hours~\cite{Yang2003,Sharova2009}. Similarly, protein lifetimes cover quite a wide range, from minutes to several days~\cite{Yen2008}. 
MiRNAs are usually stable molecules with an half-life that can span days~\cite{Rooij2007,Kai2010}, but there are cases of short-living miRNAs, as many miRNAs 
expressed in human brain \cite{Sethi2009}. Moreover, the miRNA turnover seems widely regulated as it happens for mRNAs and proteins~\cite{Chatterjee2009}. 
In summary, while the situation in which proteins are more stable than the corresponding transcripts could still be frequent, a variety of specific cases is expected. 
Therefore, we decided to take into account the mRNA dynamics and explore different regimes of molecules' half-lives. Indeed, we will show that the dynamical response of the iMSL circuit depends 
crucially on the ratio between mRNA and miRNA half-lives ($\tau_{r}/\tau{s}$).
In Figure 2A, the normalized response time $T_{ON}/T_0$ to activation is plotted as a function of the repression level   
measured as $p/p_0$, where $p_0$ is the steady-state concentration in absence of negative regulation. 
The response time of the iMSL (continuous lines) and the tSL (dashed lines) is reported for different values of the half-life ratio $\tau_{r}/\tau{s}$. 
As a first result, the iMSL can speed up the response time with a comparable efficiency with respect to their transcriptional counterpart, especially when mRNAs are degraded fastly enough. 
On the other hand, when miRNAs are short-living with respect to mRNAs, they will reach their final concentration faster than mRNAs, thus blocking more quickly the initial rise in target 
protein concentration. Therefore, the timescales of mRNA and miRNA dynamics,  determined by their half-lives, define the circuit performance in speeding up the response, as reported in Figure 2A. In Figure 2C an example of the dynamics is reported, showing an acceleration of the response for both self-regulation strategies at an intermediate level of repression. \newline
As the repression increases, the response acceleration to an activating signal relies more and more on an overshoot of protein concentration, well above the final steady state, both for iMSLs and tSLs.
  If the input signal have to drive the host gene to its functional steady-state, a large overshoot can be unwanted since it represents an unnecessary 
metabolic cost and a possible source of toxic effects~\cite{Rosenfeld2002}. Thus, there is probably a limitation in the repression strength that can be applied to minimize the 
time separation between two functional steady states. On the other hand, a regime of strong repression makes the iMSL a pulse generator, 
a feature previously associated to incoherent feed-forward 
loops~\cite{Alon2006}, which can eventually lead to adaptation as will be discussed in a following section. \newline
While the speeding up of activation is a property that iMSLs share with transcriptional incoherent feed-forward loops and tSLs, an interesting peculiarity of iMSLs emerges looking at the 
time required for $p$ concentration to reach zero, starting from a constitutive level (Figure 2D reports an example of this dynamics). 
The iMSL can  delay the switch-off kinetics of the host in the same repression regime where it can accelerate 
the activation and the extent of the introduced delay is again dependent on the mRNA to miRNA lifetime ratio (Figure 2B). This apparently counterintuitive behaviour can be easily qualitatively explained.
 When a constitutively expressed gene senses a transcription stop signal, the velocity of protein concentration decrease is established only by protein and mRNA degradation rates. For example, long living 
mRNAs are more  persistent and can be translated for a longer time after the stop of transcription, and long living proteins are obviously more resilient. 
The same is true for tSLs or transcriptional feed-forward loops: as the transcription is switched off, transcriptional repressors cannot exert any regulation and the protein level simply undergoes the exponential decrease dictated by mRNA and protein degradation. 
 On the other hand, thanks to the post-transcriptional regulation in iMSLs, for each single miRNA that is degraded the still present mRNAs sense an increase in their translation rate. 
This increase clearly depends 
on the repression  strength that miRNAs can exert (thus on the repression fold $p/p_0$) and on the relative stability of mRNAs and miRNAs ($\tau_r/\tau_s$), as a fast miRNA turnover 
leads to a higher translation rate of the remaining mRNAs. 
Eventually, the general increase in mRNA translation rate for each miRNA degradation event can 
lead to a temporary boost in protein concentration above the original steady state (see Figure 2D).\newline
It is important to notice that the dynamics just described can be altered if the miRNA acts mostly on mRNA degradation and depends on the timescale of miRNA-mRNA binding-unbinding. 
While the iMSLs can always speed up the host gene expression in activation, the delay in the the switch-off dynamics 
can vanish in case of fast miRNA-mediated induction of mRNA degradation. This issue is discussed in more details in the Additional file 1.

\subsubsection*{The circuit response dynamics can robustly keep the host gene in a high-expression state.}

In the regime of comparable mRNA and miRNA lifetimes (red curves in Figure 2) the iMSL circuit can both accelerate the response to a switch-on signal and delay the switch-off kinetics. 
This alteration of the dynamics makes the host gene ON-state (expression at maximum rate) robust with respect to a transient fading of the input activating signal, as the one depicted in Figure 3A.  
In fact, the response 
to an input fluctuation toward zero is a slow protein concentration decrease, followed by a quick recovery of the 
ON-steady-state when the fluctuation is over (Figure 3B). Only a persistent absence of signal would cause a complete disappearance of the host protein product. In this way, the cell 
could prevent a drop in concentration of physiologically necessary proteins in merely presence activator fluctuations. 
A resilient ON-state can be biologically important if it ensures the homeostatic protein level that must be robustly kept to mantain the correct phenotype or if the deactivation/reactivation is a 
costy process that have to be engaged only when undoubtedly necessary.\newline
This property can be measured more quantitatively by the distance $d$ from the ON-steady-state that is reached by the target protein level during a temporary absence of 
the input activator lasting a time $T^*$. As shown in Figure 3C,  the iMSL regulation keeps the host gene protein product close to its steady state in presence of input fluctuations that would almost switch-off a gene transcriptionally self-regulated or constitutively expressed. 

\subsection*{Intronic microRNAs, targeting their host gene, can implement adaptation and Weber's law.}

\subsubsection*{Adaptation}

Adaptation is defined as the ability of a system to respond to a change in the input but subsequently return to the original level, even if the stimulus persists. 
Adaptation is ubiquitous in signaling systems. Examples of nearly perfect adaptation range from chemotaxis in bacteria~\cite{Kollmann2005} to sensor cells in higher organisms~\cite{Matthews2003}.  
In all these systems, the benefit of adaptation can be summarized as the possibility of signal detection irrespective of the background level, thus widening the range of accessible 
signals and keeping the system robust in presence of perturbations.\newline
Simple network topologies, as negative feedback loops with a buffering node or incoherent feed-forward loops, can be at the basis of the cellular implementation of adaptation~\cite{Ma2009}. 
In this section, we investigate whether and in what conditions a post-transcriptional self-regulation 
 through intronic miRNAs can perform adaptation.\newline
It is easy to show analytically (see Additional file 1) that in the regime of strong repression ($s/h\gg1$ in the Michaelis-Menten function in Equation~\ref{reprhill}) the steady state of $p$ concentration  
is independent of the input level $q$, which is clearly a hallmark of perfect adaptation~\cite{Sontag2010}: after an eventual dynamical response to a change in $q$, 
the system always returns to its original equilibrium level. 
On the other hand, it is impossible to achieve such an independence on the input level at equilibrium using a tSL, as confirmed by the fact that circuits with 
just two molecular species are not adaptive~\cite{Ma2009}.\newline
More generally, we can evaluate the efficiency in performing adaptation giving the circuit a 
step function as input and calculating the two indexes of precision $P$ and sensitivity $S$~\cite{Francois2008,Ma2009} represented in  Figure 4A and  defined by: 

\begin{eqnarray}
P & = & \left| \frac{(p_1 -p_0)/p_0}{(q_1-q_0)/q_0} \right| ^{-1} \nonumber\\
S & = & \left| \frac{p_{max}-p_{0}}{p_0} \right| .
\label{adaptation_index}
\end{eqnarray}

$P$ is a measure of the difference in the steady-state levels before and after the stimulus, therefore it is actually an estimate of the degree of adaptation. 
Following \cite{Ma2009}, we define the minimal threshold $P>10$ to select adaptive circuits. A high value of $P$ is not enough to define adaptation since it could merely be a 
consequence of complete insensitivity to input changes. 
Thus, it is necessary to check if the peak in $p(t)$ concentration  is an effective recognizable signal. 
This condition can be formalized requiring a sensitivity $S$ above the noise level ($CV_p=\sigma_p/ \langle p \rangle$) of $p$ at steady state, as can be calculated using the 
stochastic version of the model (see Additional file 1). 
In particular, we choose the threshold $S > 2 CV_p$ (assuming a noise in the input level $CV_q=10\%$) to define a circuit ``sensitive'' to the input signal.

\subsubsection*{Weber's law}

Certain adaptive systems, besides the ability to return to their original value after a signal response, present also a degree of response that is proportional to the relative
 change in the input signal and not to its absolute value. This feature is known as Weber's law, originally introduced in the context of human sensory response. Recently, 
this dependence on input fold-change was demonstrated experimentally in eukaryotic signaling systems~\cite{Goentoro2009b,Cohen-Saidon2009}, and theoretically the feed-forward loop 
topology was proposed as a candidate to Weber's law implementation in gene regulatory networks~\cite{Goentoro2009}.\newline
Once again, it is natural to examine in what conditions also the minimal iMSL circuit can satisfy Weber's law.\newline
It is possible to show analytically (see Additional file 1) that iMSLs are responsive to input fold-change if three conditions are satisfied:

\begin{itemize}
\item Strong repression: $s/h\gg1 \Rightarrow k_p(s) \approx k_p h / s$ (condition for perfect adaptation),
\item Almost linear promoter activation  $k_r(q) \approx  q k_r/h_r$,
\item Fast mRNA dynamics (short mRNA half-life with respect to miRNA and protein ones): $r(t) \rightarrow r_{ss}$.
\end{itemize}

As for the case of adaptation, we can quantify the efficiency in Weber's law implementation for a generic set of biochemical parameters. 
To this aim, a two step input function is provided such that each step has the same fold-change but different background levels (see Figure 4B). As previously proposed~\cite{Goentoro2009}, 
the error $E$ in recognition of fold changes can be quantified using the difference in the response peaks: 

\begin{equation}
E = \left| \frac{p_{max2}-p_{max1}}{p_{max1}} \right|.
\label{weber_index}
\end{equation}

\subsubsection*{Parameter space of adaptation and Weber's law}

Using the observables defined in Equations~\ref{adaptation_index} and~\ref{weber_index}, it is possible to explore 
the conditions in which adaptation and Weber's law are successfully performed by iMSLs.
An illustrative example is depicted in Figure 4C, where two effective parameters are varied: the effective promoter activation $q/h_r$, and 
$1/h$ which  measures the repression strength since $h$ is the number of miRNAs necessary to reduce to one half the target translation rate. 
The grey region depicts the parameter space where precise adaptation is performed ($P>10$),  
while in the excluded red region the dynamical response of the circuit is not able to go beyond the noise level ($ S<2CV_p$). The $E$ value is reported with the color code in the legend when the minimal condition $E<0.1$ is satisfied, i.e the two steps of the input  
produce the same response within $10\%$.  
Adaptation and Weber's law can be encoded by iMSLs in a parameter region that span several orders of magnitude of the effective parameters. Therefore, the only constraint is that the effective 
parameters have to approach the appropriate limits, without the need of fine-tuning.\newline
It should be noticed that the general condition of strong repression required for both functions is limited by the circuit sensitivity.
This is partially due to the fact that a too strong repression can rise the noise level of the circuit (see next section) making 
the achievement of a signal significantly above fluctuations harder.\newline
It is interesting to consider what are the functional advantages that these two functions can provide to the host gene. 
Both adaptation and Weber's law can bestow robustness to the expression program of the host gene. 
An expression state that is not influenced by constant inputs thanks to adaptation is robust with respect to the 
ubiquitous cell-to-cell variability in TF concentrations, but it is still responsive to signals that induce dynamic variations of TF levels. 
When additionally Weber's law is implemented, also the dynamic response can be kept homogeneous 
in a cell population. In fact, in this case the response profile is only due to the input fold-change and not on its absolute value that is affected by the potentially 
variable background level~\cite{Goentoro2009}. 
Moreover, Weber's law naturally encodes a noise filter. In fact, since the noise level is expected to scale with the background TF concentration,  
a dependence on fold-change can naturally rescale appropriately the threshold at which the response is triggered, thus allowing a better signal/noise discrimination in different background conditions~\cite{Goentoro2009}.

\subsection*{Autoregulation via intronic microRNAs reduces the host gene expression fluctuations.}

All the functions of iMSLs  discussed so far contribute to enhance the robustness of the host gene expression. It is therefore natural to analyze a stochastic model 
of iMSLs to test directly their ability to filter out fluctuations.    
The stochastic analysis of the system is reported in detail in Additional file 1. 
The results in terms of noise-buffering properties at the steady state for the iMSL are similar to those obtained for the 
incoherent miRNA-mediated feedforward loops (see~\cite{Osella2011}). 
By filtering fluctuations propagating from the upstream TF, the steady-state target protein level achieved with an iMSL is less noisy 
than the same target amount obtained with a simple sTF or a tSL (Figure 5A-B). 
In particular, the target noise  $CV_p$ for the iMSL shows a U-shaped profile with a well defined minimum, thus allowing us to identify the 
parameter values that optimize the noise reduction properties (Figure 5B). 
This prediction could in principle be tested tuning the repression strength, as shown in~\cite{Dublanche2006} for a tSL. 
Also a tSL can in fact optimally filter noise for well defined values of repression strength~\cite{Shahrezaei2008,Dublanche2006,Singh2009}, as shown in Figure 5B (orange dots and line). 
For this circuit the mechanism is well understood: an excessive increase of the repression strength (while potentially improving the noise reduction of the circuit) 
reduces the copy number of mRNAs and proteins with a consequent 
rise in intrinsic fluctuations (which can overcome attenuation). Thus, there is just a well defined range of repression strength for which the noise reduction is optimal, 
as shown in experiments~\cite{Dublanche2006}.\newline
It is interesting to notice that, even if iMSLs and tSLs show similar noise reduction properties, 
the miRNA-mediated self-regulation actually performs better than the transcriptional self-regulation. 
As it is possible to see in Figure 5A (where histograms and continuous lines are respectively the
result of Gillespie's simulation with full nonlinear dynamics and gamma distributions with analytically calculated moments), the probability distributions of the 
target protein level for the three circuits are different. 
Both autoregulatory circuits lead to a target distribution less sparse than a sTF, showing that they effectively reduce fluctuations, but the iMSL distribution is clearly more peaked than the tSL one. 

Similarly, both  self-regulation strategies show an optimal noise buffering for an intermediate repression strength, but again the 
attenuation is larger in the miRNA-mediated case (see Figure 5B). 
This is more clearly shown in Figure 5C-D, where the noise reduction $CV_{p}/CV_{p}^{sTF}$ (with $CV_{p}^{sTF}$ representing the target noise in the case of a simple transcription 
unit producing the same mean amount of proteins) is reported for the two autoregulatory circuits.  Noise reduction is explored for different levels of  transcriptional activation ($\langle q \rangle /h_r$)  
and target repression ($\langle p\rangle/\langle p_0 \rangle$, where $\langle p_0 \rangle$ is the target mean value in absence of repression) to shed light into noise control and target suppression interdependence.
In the regime where the target is more sensitive to TF fluctuations, i.e. $q$ is far from saturating the promoter, the iMSL can reduce the fluctuations up to a factor $0.5$ (Figure 5C), 
while the tSL (Figure 5D) is much less effective. 
Moreover, the heat maps in Figures 5C-D indicates that iMSL can buffer fluctuations over a wider range of conditions as well as to a greater extent.
\newline
As pointed out in~\cite{Osella2011}, an optimal miRNA-mediated noise buffering does not necessarily require a strong
repression. Indeed, Figure 5C shows that a reduction of the mean protein expression to $50\%$ of its constitutive level is sufficient to reduce the noise by approximately $40\%$. 
This means that the intronic miRNA can keep the expression of its host gene in its homeostatic regime,  
while filtering out fluctuations, without exerting a strong reduction of its concentration. This result agrees well with the observation that miRNAs act often to fine-tune 
their targets rather then to switch them off completely~\cite{Baek2008}.

\subsection*{Sketch of the one-to-many topology-function map}

This section summarizes the functions found to be associated to intronic miRNA-mediated self-loops into a qualitative ``map of functions", 
showing the different, although overlapping, ranges of biochemical parameters in which each specific function is optimized. 
The emerging map between parameter values and functions can be useful to understand the presence of the iMSL architecture in different 
biological contexts and gives general guidelines for the 
design of synthetic circuits with a desired behaviour, well beyond the simple suggestion of a network topology. \newline
As Figure 6A shows, strong repression ($\langle p\rangle /\langle p_0 \rangle \ll 1 $) is a general requirement for the implementation of adaptation and Weber's law,  the latter additionally requiring an almost linear activation of transcription ($\langle q\rangle / h_r \ll 1$). A sufficienlty strong repression is also required to confer robustness to  
the  high-expression state (induced by strong activation $\langle q\rangle / h_r \gg 1$) of the host gene in presence of input temporary drops. 
On the other hand, for intermediate host activation, where the host gene promoter is highly sensitive to changes in the TF concentration, the iMSL can  efficiently buffer fluctuations 
at steady state without the need of strong repression. \newline
Looking at a finer scale the strong repression regime ($\langle p\rangle /\langle p_0 \rangle < 1/2 $), a smooth transition in the dynamical behaviour of the circuit can be observed (see Figure 6B). 
At first, the host gene is able to fastly transit between two well distinct steady states after induction. When the repression is further increased, 
this fast ON-activation relies increasingly on an overshoot well above the final equilibrium at which the dynamics asympotically relaxes. Therefore, the concentration profile resembles a pulse. Finally, 
for high enough repression the system returns to the initial steady state after the pulse, a necessary condition for the implementation of adaptation and Weber's law.
\newline   
The relative half-life of the molecules involved, in particular of miRNAs and mRNAs, is another ingredient that can strongly influence the dynamic behaviour (see Figure 6C). For example, a miRNA half-life comparable to the mRNA 
one allows a trade-off between 
acceleration of the ON-dynamics and delay of the OFF-dynamics, making the state of high expression of the host gene robust to fluctuations. On the other hand, mRNA lifetime 
must be short with respect to the other molecules lifetimes for a dynamical response following Weber's law.\newline 
Autoregulation via intronic miRNAs shares structural similarities with miRNA-mediated incoherent feed-forward loops, 
that represent a diffused and functionally relevant motif in regulatory networks~\cite{Tsang2007,Re2009,ElBaroudi2011,Eduati2012,Osella2011}. 
In fact, iMSLs can be thought as a mimimal feed-forwad topology with perfect co-expression of the target gene and the miRNA buffering node.   
Therefore, the findings presented here could be easily generalized to miRNA-mediated incoherent feedforward loops 
adding new pieces to our understanding of microRNA regulation in simple circuits.\newline
Finally, the present analysis of the iMSL functions considers the circuit as isolated, while realistically a single microRNA can target hundreds of genes. As recently pointed out, 
the degree of repression of a target depends on the level of expression of all possible target genes~\cite{Levine2007,Arvey2010}, since their mRNAs can dilute the pool of available miRNAs. 
Therefore, the expression profile of alternative miRNA targets is a variable that can potentially alter the dynamics of iMSLs (as shown for incoherent feed-forward loops~\cite{Osella2011}), and thus have to be carefully taken into account in experimental tests on endogenous iMSLs.

\subsection*{Identification of intronic miRNA-mediated self-loops in human.}

In this section, we briefly describe our bioinformatic search of iMSLs. Our main goal is to provide an updated list of candidates to 
eventually test our theoretical predictions.
We performed a genome wide search of intronic miRNA-mediated self-loops along the lines of two papers which recently addressed the same issue \cite{Megraw2009, Hinske2010}. 
The differences between our results and those quoted in \cite{Megraw2009, Hinske2010} are mainly due to a different choice of
the algorithms used to predict miRNA targets and in some cases 
to the use of updated versions of the corresponding databases.
We identified the same strand intronic miRNAs using as reference the Ensembl (release 57) database. A summary of our results is reported in Table 1S of Additional file 1 and in Figure 7A, 
where the percentage of intergenic versus intragenic miRNAs is plotted and, for the intragenic ones, the relative ratio of exonic versus intronic miRNAs and of same-strand versus 
opposite strand is also reported. 
Target identification was performed using 8 different algorithms: 
TargetScan human v. 5.0 \cite{Lewis2003, Friedman2009}, miRanda - release 2008 \cite{John2004, Betel2008}, RNA22 \cite{Miranda2006}, PITA-4way \cite{Kertesz2007}, MirTarget2 \cite{Wang2008}, 
PicTar \cite{Krek2005}, Diana microT v.3 \cite{Maragkakis2009} and TargetMiner v.1 \cite{Bandyopadhyay2009}. 
In this respects, our analysis could be considered as a combination and 
extension of the one reported in reference~\cite{Hinske2010} (where only the first 6 algorithms were used) and the study of \cite{Megraw2009} (where the authors used their own prediction algorithm). 
We were able in this way to find a total of 77 iMSLs confirmed by at least one algorithm (details are reported in Table 2S of Additional file 1). Since these algorithms are very different, we did not try to 
give an absolute score to our results, but ordered them starting from those which were assessed by the largest number of target prediction methods (Figure 7B and Table 2S of Additional file 1). 
Following a standard recipe (see \cite{Hinske2010} for a similar choice) we consider the number of different algorithms that agree on a certain target prediction as a measure of the 
confidence of such prediction. Interestingly, 28 of our iMSL agree with previous predictions of iMSLs \cite{Megraw2009} and for
 two of them an experimental validation of the miRNA-host 
gene regulation exist \cite{Megraw2009,Sun2010,Nikolic2010}. 
Moreover, a recent study \cite{Ma2011} provides evidence supporting a feedback mechanism between miR-438 and IGF2, in agreement with our list of iMSLs predicted by only one method.
In order to test if these iMSLs are over-represented we performed two independent enrichment tests. First we performed a reshuffling of the host genes while keeping miRNA target 
predictions unchanged. Second, we randomized the union of the datasets of miRNA target
predictions obtained with the eight algorithms discussed above, keeping the host genes unchanged.  In both cases we evaluated the $Z$ score which turned out to be  $Z = 4.63$ for the first 
test and  $Z = 5.52$ for the second one. The results of the tests are plotted in Figure 7C. They suggest, in agreement with what already observed in \cite{Ozsolak2008, Megraw2009}, 
that this particular class of network motifs is under positive selection.

\section*{Conclusions}

This study presents a fairly comprehensive survey of the possibile functions associated to a miRNA-mediated circuit composed of one protein-coding gene (host gene) negatively regulated by a miRNA located in one of its introns. 
In particular, we have shown that, thanks to the miRNA-mediated self-regulation, the host gene expression responds to input changes with an altered 
timing, its response can be adaptative and follow Weber's law, and fluctuations propagating from the upstream network are buffered at steady state.
Each of these functions can confer robustness to the expression program of the host gene, suggesting that miRNA-mediated self-loops represents a simple 
homeostatic control. For example, adaptation makes the host gene expression level at equilibrium independent of the cell-to-cell variability of 
transcription factor expression without compromising its sensitivity to input changes. Similarly, the host expression dynamics, as modified by miRNA autorepression,  
can mantain the host gene in a high-expression state in the face of downwards fluctuations in activators' concentration. 
The association of miRNA-mediated self-loops with functions with different specificities, but apparently the same final aim, suggests that, depending on 
the desired level of the host gene expression and on the type of fluctuations that have to be more frequently filtered out, the details of the 
regulatory interactions and the characteristics of the molecules 
involved could have been fine-tuned over evolutionary timescales accordingly. Such a fine-tuning of expression parameters has been shown to be possible even over short timescales in \textit{in vitro} evolutionary experiments~\cite{Dekel2005}.\newline
The comparison with an unregulated transcriptional unit and with a transcriptional negative feedback indicates that the specificities of miRNA regulation 
makes the post-transcriptional circuit better suited to implement a homeostatic control. This result is in line with the accumulating clues that miRNAs 
 can help cells to function reliably in presence of perturbations~\cite{Inui2010,Li2009,Li2006,Ebert2012}. 
 \newline  
Finally, our systematic analysis of the constraints on biochemical parameters necessary to optimize each function can guide the realization of synthetic versions of miRNA-mediated self-loops, as well as contribute to the understanding of the role of their many occurrences in endogenous networks. In this perspective, we also provide a list of bioinformatically predicted miRNA-mediated self-loops in human for future experimental tests.

\section*{Methods}

\subsection*{Stochastic simulations}
Simulations were implemented by using Gillespie's first reaction algorithm \cite{Gillespie1976}. 
The reactions simulated are those presented in Figure 1 with additional transcription, translation and degradations for the input transcription factor $q$. Reactions that depend on a regulator were allowed to have as rates the corresponding full nonlinear functions. Results in Figure 5 are at steady state, which is assumed to be reached when the deterministic evolution of the system in analysis is at a distance from the steady state (its asymptotic value) smaller than its 0.05\% (more than 10 times the protein half-life). Each data point or histogram is the result of 100000 trials.

\subsection*{Bioinformatic methods}

In order to identify human intragenic miRNAs, and associate them to their host genes, we used Ensembl-release 57 database. 
We collected the data of human known protein coding genes (Total Ensembl Gene Identifiers (ENSG) entries 22.257, 
for each one we considered the longest Transcript Identifiers (ENST)).
The data on human miRNAs  were extracted from Ensembl v.57, that includes miRBase v.13 (Table 1S). 
To identify the iMSLs, we used eight tools for miRNA/target gene interaction 
predictions: TargetScan human v. 5.0 ~\cite{Lewis2003, Friedman2009}, miRanda - release 2008~\cite{John2004, Betel2008}, RNA22~\cite{Miranda2006}, PITA-4way~\cite{Kertesz2007}, 
MirTarget2~\cite{Wang2008}, PicTar \cite{Krek2005}, Diana microT v.3  \cite{Maragkakis2009} and TargetMiner v.1 \cite{Bandyopadhyay2009}. 
To test the over-representation of the putative iMSLs, we performed two different types of randomization strategies. 
Specifically, we randomly permuted 1000 times the intronic miRNA / host-gene link and the union of miRNA/target gene 
interactions datasets predicted by the different algorithms. Then we used these 1000 randomized sets 
to rebuild a list of the iMSLs. To show the statistical significance of our results we calculated the Z-score (Z) for each randomization.

\bigskip

\section*{Competing interests}
The authors declare that they have no competing interests.

\section*{Author's contributions}
CB, MO and MC designed research. CB, MO, ME, DC performed research.  CB, MO and MC wrote the paper. All authors read and approved the final manuscript.

\section*{Acknowledgements}
  \ifthenelse{\boolean{publ}}{\small}{}
We would like to thank Marco Cosentino Lagomarsino for useful comments on the manuscript.
MO acknowledges support from the Human Frontier Science Program Organization (Grant RGY0069/2009-C).


\newpage
{\ifthenelse{\boolean{publ}}{\footnotesize}{\small}
 \bibliographystyle{bmc_article}  
  \bibliography{arXiv_selfloop.bib} }     


\ifthenelse{\boolean{publ}}{\end{multicols}}{}

\begin{figure}[h!!]
\includegraphics[scale=0.8]{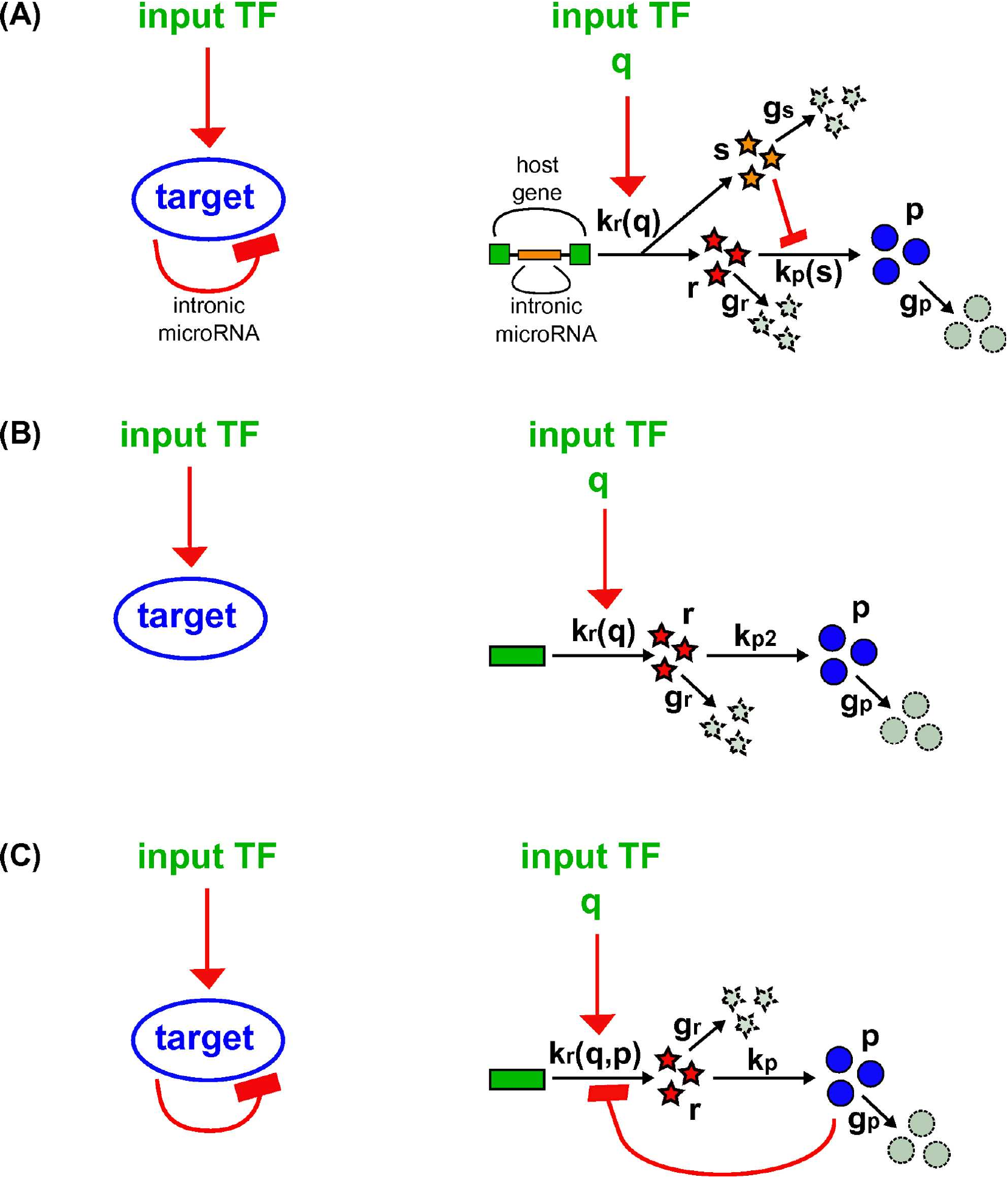}
\centering
\caption{ {\bf Representation of iMSL and the two circuits used for comparison.}
      Schematic views of (A) an intronic miRNA-mediated self-loop (iMSL); (B) a gene simply activated by a TF (sTF); (C) a  transcriptional self-regulation (tSL).
A more detailed representation of the three circuits is on the right of the figure. 
Green rectangles are DNA-genes; $s$ and $r$ are the transcribed miRNAs and mRNAs (orange and red stars respectively) which can eventually be degraded (broken grey stars). 
mRNAs can be translated into proteins $p$ (blue circles) and proteins can be degraded (broken grey circles). 
The reaction rates are reported along the corresponding black arrows: $k_{r}(q)$ and $k_{r}(q,p)$ for transcription; $k_{p}(s)$, $k_{p2}$ and $k_p$ for translation;  
$g_s$, $g_r$ and $g_p$ for degradation. Red arrows represent activations, while red lines ending in bars are repressions.
For the iMSL and the sTF, the transcription rates are functions of the amount of TFs $q$, while for the tSL the transcription rate is also a function of the target 
protein $p$. In the iMSL, miRNA regulation makes the rate of translation a function of the amount of miRNAs $s$.}
\end{figure}

\begin{figure}[h!!]
\includegraphics[width=\textwidth]{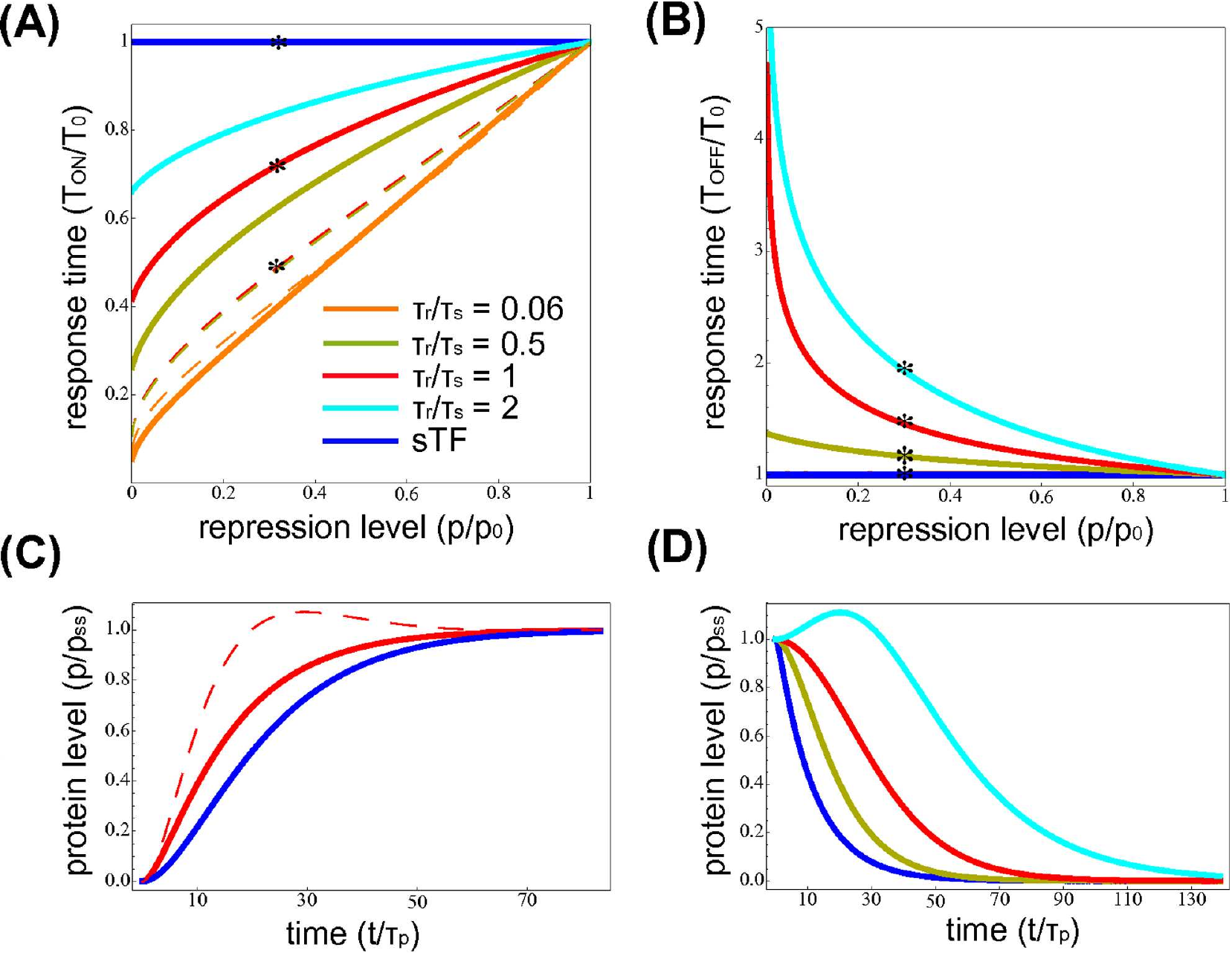}
\centering
\caption{{\bf Autoregulation via intronic miRNAs speeds up the host gene activation and delays its deactivation. }
      (A) Activation response time: the response time $T_{ON}$, normalized by the response time of the sTF $T_0$, is plotted as a function of the repression level 
$p/p_0$ (e.g. $p/p_0 =1$ means no repression) for different values of mRNA/miRNA lifetimes. 
Both  the iMSL and the tSL (continuous and dashed lines respectively) are able to accelerate 
the target response time with respect to the sTF (blue horizontal line). 
Each color corresponds to different values of mRNA and miRNA relative stability ($\tau_{r}/\tau_{s}$). In particular, in this plot   
 the degradation rate of mRNAs ($g_r$) and proteins ($g_p$) are fixed,  while different values of miRNA degradation rate ($g_s$) give the different curves. 
The half-life ratio $\tau_{r}/\tau_{s}$ affects in principle also the tSL dynamics, since it is constrained to have same final mean protein concentration, 
but actually the dependence is weak and the corresponding curves tend to collapse.   
(B) Deactivation response times: the response time $T_{OFF}$, normalized by the response time of the sTF $T_0$, is shown for different repression levels. 
The blue line corresponds to the sTF (reference time), while the response time for the iMSL is plotted for different miRNA half-lives (same color code of A). 
The iMSL induces a delayed host gene response for the same values of repression that consent an acceleration of the activation response. 
(C) Example of the temporal evolution of the target protein concentration in activation for the three regulatory strategies. The parameter values correspond 
to the stars in the above plot A and time is in protein half-life units.  
(D) Example of the deactivation dynamics of the target protein concentration in the iMSL case for different mRNA/miRNA half-lives (corresponding to the stars in the above plot B). 
The parameter setting for this panel is the following: protein half-life $\tau_p = 8$ hours, mRNA half-life $\tau_r = 30$ minutes, $h=1000$,  $k_r=0.212819~s^-1$, $k_p=0.0048~s^-1$.}
\end{figure}

\begin{figure}[h!]
\centering
\includegraphics[scale=0.9]{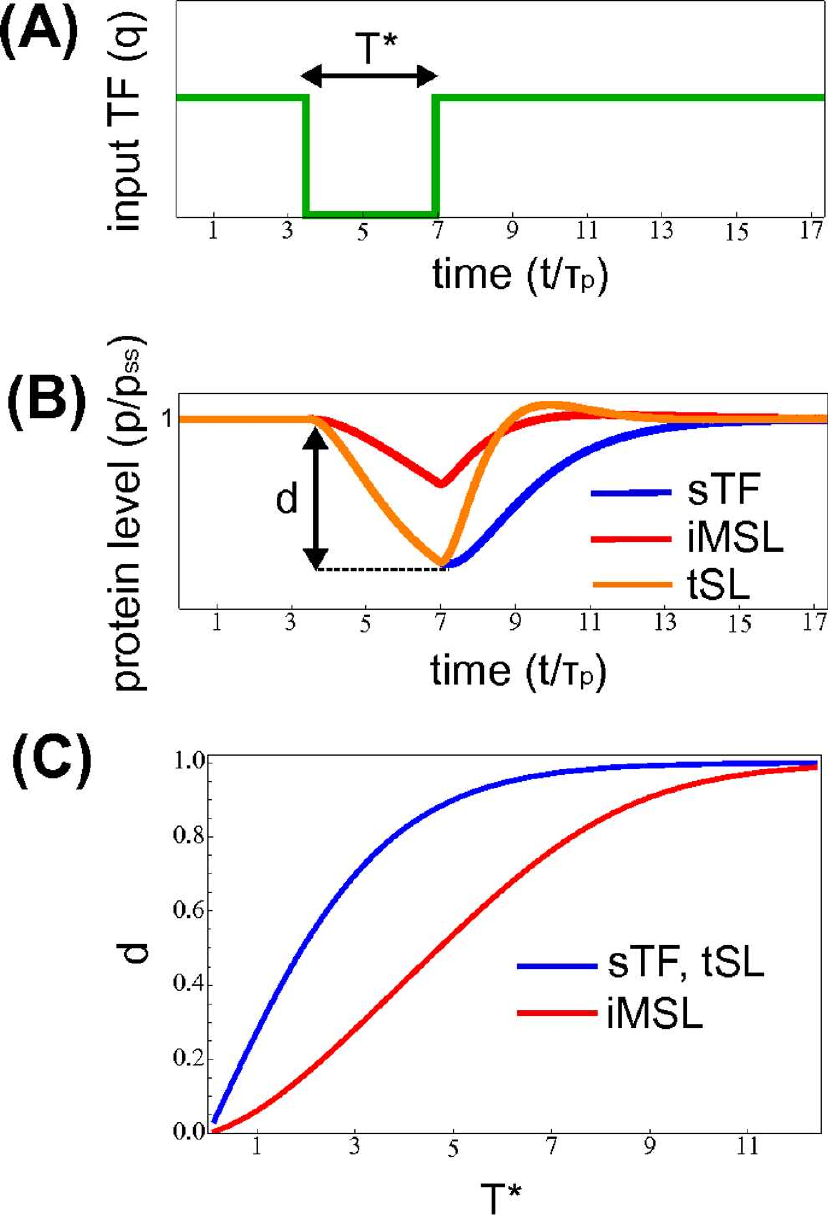}
  \caption{{\bf  miRNA-mediated self-loops can keep the host gene expression robustly in a ON-state.}
      (A) Schematic representation of a transient drop of the input TF $q$  of duration $T^{*}$ (time is in protein half-life units $\tau_p$)
(B) Response of the three circuits to the temporary absence of signal depicted in A. 
The iMSL response (red line) is a slow protein concentration decrease, followed by a quick recovery of the 
ON-steady-state when the fluctuation is over. For a tSL (orange line) or a sTF (blue line) the switch-off dynamics is just due to mRNA and protein degradation. 
Even if the transcriptional negative feedback can accelerate the recovery, the distance 
$d$ from the ON steady state that is reached 
by the target protein level during the temporary absence of the input activator is determined by the switch-off response.  
(C) The distance $d$ from the ON steady-state reached by the target level is plotted as a function of the duration of the TF $q$ absence ($d=1$ when 
the protein level  reaches zero). The iMSL circuit (red line) requires a more persistent absence of signal to show a significative reduction of the host protein product level 
with respect to the tSL or the sTF (blue line). The parameter values are the same of Figure 2, with comparable mRNA and miRNA stability ($\tau_{r}/\tau_{s}=1$).}
\end{figure}

\begin{figure}[h!]
\centering
\includegraphics[scale=0.8]{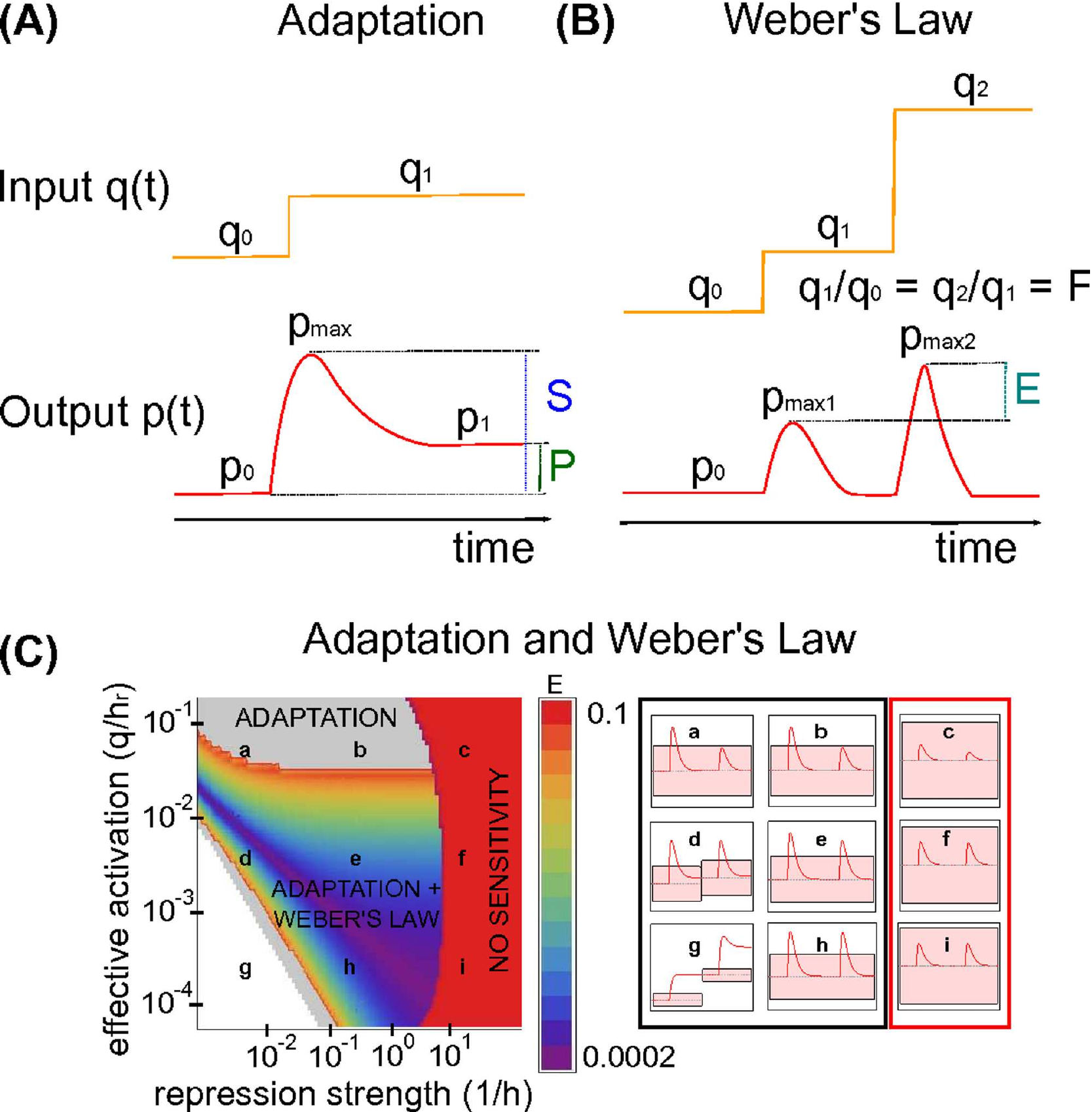}
\caption{{\bf  Adaptation and Weber's law implementation via intronic miRNA-mediated self-regulation.}
      (A) Schematic view of adaptative behaviour for a step-like input. $S$ and $P$ are the sensitivity and precision measures described in the main text.
(B) Schematic view of Weber's law implementation for a two-step input function. $E$ is the error in fold-change detection, as defined in the main text.     
(C) A summarizing heat-map of the imLS performances in implementing adaptation and Weber's law, as a function of the effective activation $q/h_r$ and the repression strength $1/h$. 
The grey region is the adaptive region ($P>10$ and $S>2CV_p$), while the region where the system implements also Weber's law ($E<0.1$) is depicted with a color code representing the $E$ 
value as reported in the legend. In the red zone the system is not sensitive enough to input variations ($S<2CV_p$). 
On the right, the target protein-level response to a two-step input is reported for the parameters values identified by the corresponding lower-case letters in the heat-map. 
The shaded regions correspond to the $2CV_p$ sensitivity threshold, showing that for a too strong repression the circuit response cannot produce a signal beyond the noise level 
(plots in the red rectangle).  
The parameter setting is the following: mRNA and protein half-lifes as in Figure 2, miRNA half-life is $\tau_s=8$ hours, $k_r=2.12819~s^-1$, $k_p=0.048 ~s^-1$, 
the input function starts from an initial value of $q_0=40$ and makes two consecutive steps with fold-change $F=4$.}	
\end{figure} 

\begin{figure}[h!!!!]
\centering
\includegraphics[scale=0.9]{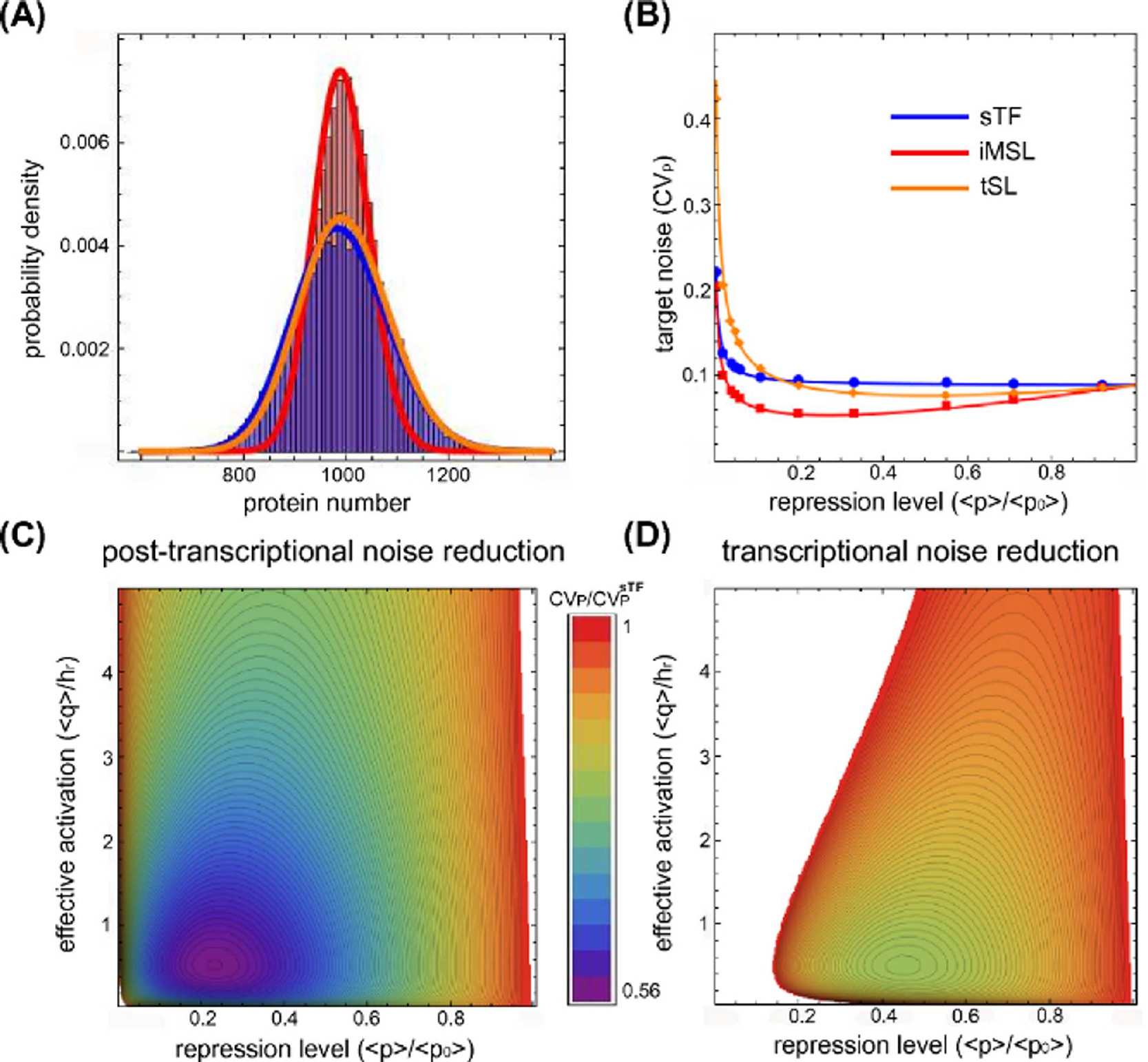}
 \caption{{\bf  Intronic miRNAs can buffer noise in host gene expression.}
(A) An example of the target protein distributions for the three circuits (repression level $\langle p \rangle /\langle p_0 \rangle =0.2$). Lines are gamma distributions with 
first two moments calculated analytically,  while histograms are the result of Gillespie simulations. 
The distribution for the iMSL circuit (red line and histogram) is the narrowest, showing that, even if also a tSL (orange line and histogram) can reduce noise with respect to a sTF 
(blue line and histogram), the iMSL is outperforming.
(B) Target noise $CV_p$ as a function of the repression strength $\langle p \rangle / \langle p_0 \rangle$ for the three circuits. Lines are analytical predictions, while dots are the result of 
Gillespie simulations. Given a noise level $CV_q \simeq 0.2$ in the upstream transcription factor, both the iMSL (red lines and dots)  and the tSL (orange line and dots) shows a minimum 
of noise reduction with respect to the sTF (blue line and dots), but the level of fluctuations in the iMSL case is clearly lower. 
(C,D) Noise reduction on the target protein level achieved by the iMSL and the tSL respectively. The noise reduction $CV_p/CV_{p}^{sTF}$ (where $CV_{p_0}^{sTF}$ measures the fluctuations around the 
same mean level for a sTF)  is evaluated at different degrees of transcriptional activation $\langle q \rangle /h_r$ and repression $\langle p\rangle / \langle p_0\rangle $. 
The same color gradient is used in both heat maps, showing that the iMSL reduces  fluctuations on a larger parameter region and to a greater extent. In the white regions $CV_p > CV_{p}^{sTF}$.  
The parameter values are the following: mRNA half-life $\tau_r =\tau_w = 30$ minutes, protein half-life $\tau_p=\tau_q= 1.5$ hours, 
$k_w= 3.4~ 10^{-3} s^{-1}$, $k_q= 8.7 ~10^{-3} s^{-1}$, $k_r=0.155$, $k_p=4.8 ~10^{-3} s^{-1}$. }
\end{figure}

\begin{figure}[h!]
\centering
\includegraphics[scale=0.68]{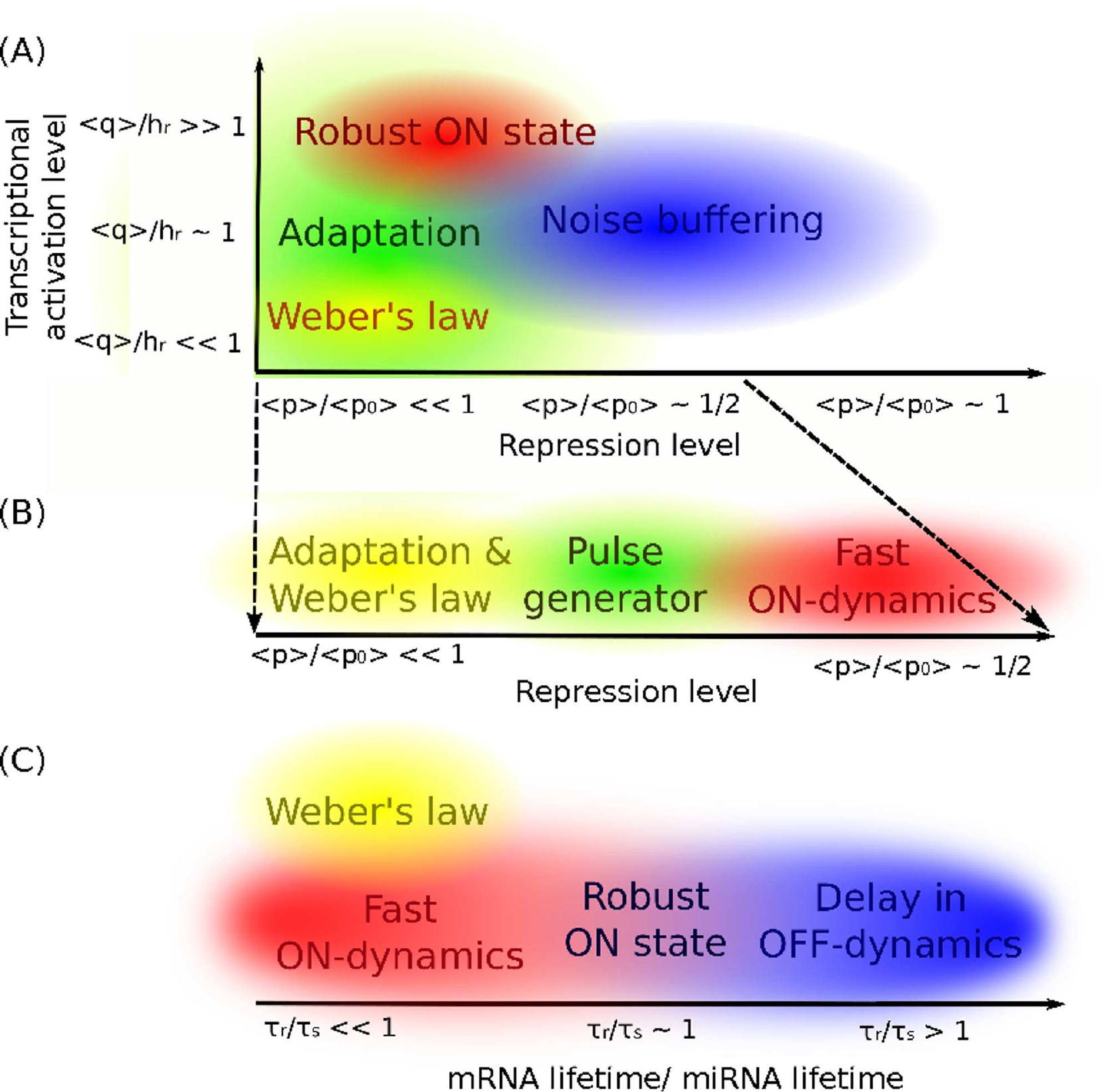}
\caption{{\bf  Map of functions for an intronic miRNA-mediated self-loop.}
(A) An ON-state of host gene expression is defined by full promoter induction ($\langle q \rangle /h_r \gg 1$), 
and a sufficiently strong miRNA repression ($\langle p \rangle / \langle p_0\rangle < 0.5$) can keep it robust in presence of temporary drops in the activator concentration. 
In the strong repression regime ($\langle p \rangle / \langle p_0 \rangle \ll 1$) adaptation can be observed, and for almost linear transcriptional activation 
($\langle q \rangle /h_r \ll 1$) the host gene response can show an adaptive dynamics following Weber's law. 
Fluctuations can propagate from the upstream TF more efficiently if the target promoter is highly sensitive to changes in TF level ($ \langle q \rangle /h_r \approx 1$), 
thus in this parameter region noise buffering is more relevant, with a maximum in efficiency for intermediate repression ($ \langle p \rangle / \langle p_0\rangle  \approx 0.3$). 
(B) A zoom on the strong repression region shows a transition between different dynamics. A step input can induce a fast transition of the host gene expression between 
two distinct steady-states, but increasing further the repression the two steady states become progressively closer, up to their overlap when adaptation and Weber's law are implemented. 
(C) The dynamics is strongly influenced by the relative stability of miRNAs and mRNAs. A short mRNA lifetime is a condition for Weber's law implementation and contributes to 
the fast switch-on of the host gene expression. On the other hand, the delay in the switch-off dynamics is larger for short-living miRNAs. In the intermediate region, 
where the two half-lives are comparable, the trade-off between the two dynamical properties makes the highly-expressed state of the host gene robust with respect to fluctuations in the activator. }
\end{figure}

\begin{figure}[h!]
\centering
\includegraphics[width=\textwidth]{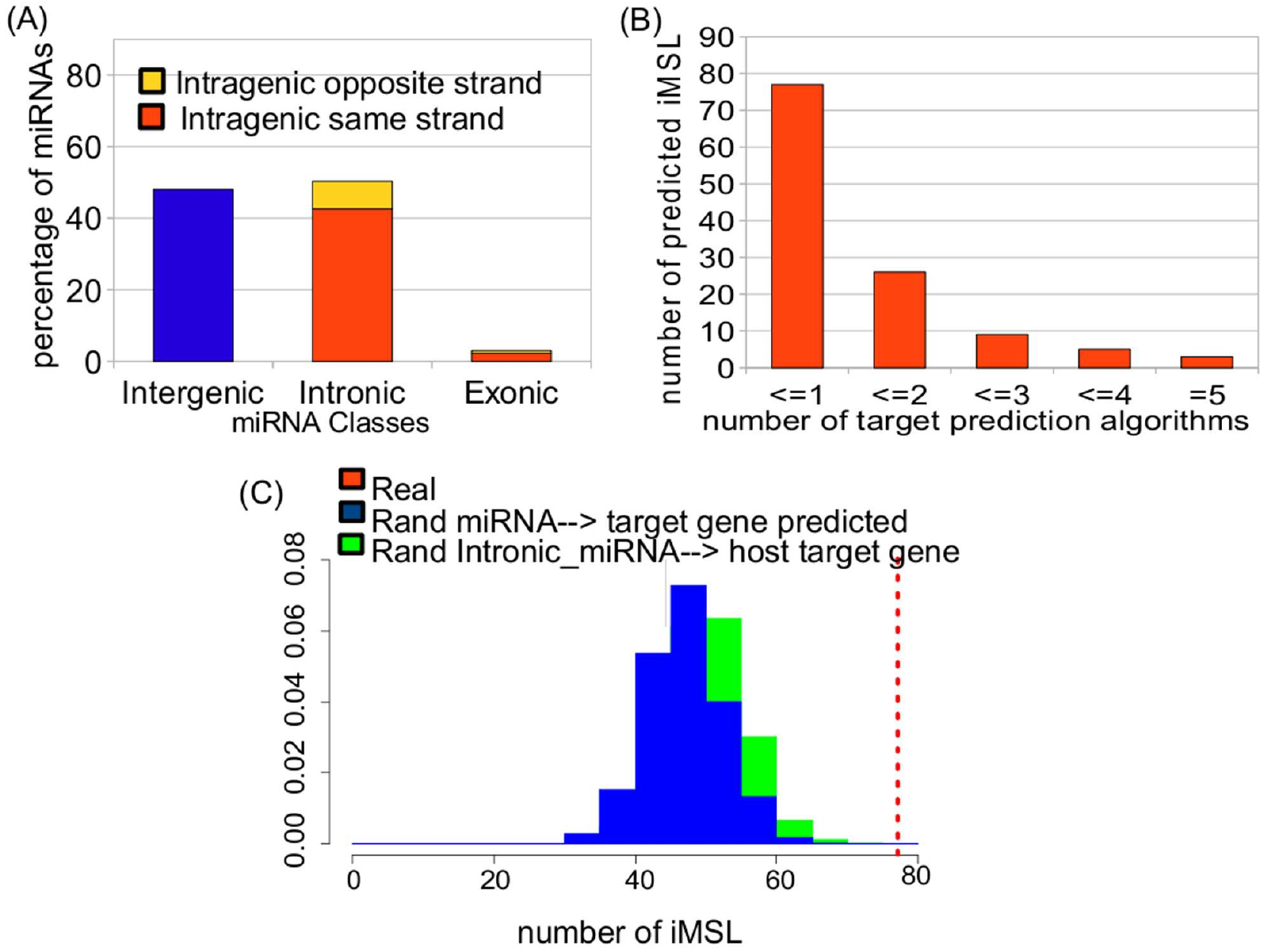}
  \caption{{\bf  Classification of miRNAs and randomization results.}
(A) Classification of human miRNAs based on the percentage of intergenic and intragenic miRNAs. Intragenic miRNAs are divided in exonic and intronic miRNAs and each group can be further classified as same strand or opposite strand. 
All UTR miRNAs were included in the group of the exonic miRNAs. Data are reported in Table 1S of Additional file 1. 
(B) Number of intronic miRNA-mediated self-loops as a function of the number of target prediction methods in agreement: 
we found a total of 77 iMSLs predicted by at least one prediction methods, 25 of them are predicted by at least 
two different methods and only three of them are predicted by 5 different methods. 
(C) Results of the permutation test: the number of  iMSLs in the human network is plotted as a dashed line alongside the distributions (normalized histograms) of the number of iMSLs found  
using the two randomization strategies (described in the main text) over 1000 experiment repetitions. }
\end{figure}

\clearpage




\ifthenelse{\boolean{publ}}{\end{multicols}}{}

\newpage
\clearpage


\setcounter{page}{1}
\setcounter{equation}{0}
\setcounter{figure}{0}

\title{Gene autoregulation via intronic microRNAs and its functions - Supplementary Information}
 

\author{Carla Bosia\correspondingauthor$^{1,2,\dagger}$ %
       \and
         Matteo Osella\correspondingauthor$^{3,4,\dagger }$ %
	\and
	  Mariama El Baroudi$^{5}$ %
	 \and
	  Davide Cor\'{a}$^{2,6}$ %
	\and 
 	Michele Caselle$^{2,7}$ %
}


\address{%
     \iid(1) Human Genetic Foundation, Molecular Biotechnology Center, University of Torino, V. Nizza 52 a, I-10126, Torino, Italy.  
     \iid(2)  Center for Complex Systems in Molecular Biology and Medicine, University of Torino, V. Accademia Albertina 13, I-10100, Torino, Italy.    
	\iid(3) Genomic Physics Group, FRE 3214 CNRS “Microorganism Genomics”, France.
\iid(4) Universit\'{e} Pierre et Marie Curie, 15 rue de L'\'{E}cole de M\'{e}decine, Paris, France.
\iid(5) LISM (Laboratory for Integrative System Medicine), CNR,  Via G. Moruzzi 1, Pisa, Italy.
\iid(6) Systems Biology Lab, Institute for Cancer Research and Treatment (IRCC), School of Medicine, University of Torino, Str. Prov. 142, Km. 3.95, Candiolo I-10060, Torino, Italy.
 \iid(7) Dipartimento di Fisica Teorica and INFN, University of Torino, V. Pietro Giuria 1, I-10125, Torino, Italy.\\
$\dagger$ Equal contributors
}%

\maketitle

\tableofcontents

\newpage

\section{Mean field analysis of the dynamics}

In this section the deterministic description of the regulatory circuits in analysis is reported in more details. On this description is based the evaluation of the response times presented in the main text.

\subsection{Simple transcriptional unit (sTF)}

We first consider the dynamics of a simple transcription unit (scheme in Figure 1B of the main text), for which the time evolution can be evaluated analytically in the two cases of interest:  the dynamics of the switch-on and switch-off processes.\newline 
The system of equations representing the sTF dynamics is:
\bea
\label{det-sing}
 \frac{dr}{dt} &=& k_r(q) -  g_r r  \nonumber \\
 \frac{dp}{dt} &= &  k_{p} r-  g_p p ,
\eea

where $k_{r}(q)$ is the nonlinear increasing function of the TF concentration $q$ reported in the main text in equation 1.\newline
With the target promoter exposed to full activation ($q/h_r \gg 1$), the transcription rate reduces to $k_r(q)\simeq k_r $ 
and it is possible to calculate 
how the final steady state is approached by the various molecular species starting from the initial condition $p(0)=r(0)=0$:

\bea
\label{on}
\frac{r(t)}{r_{ss}} &=& (1 - e^{-g_r t}) \nonumber \\
\frac{p(t)}{p_{ss}} &=& \frac{g_p (1 - e^{-g_r t}) - g_r (1 - e^{-g_p t})}{g_p - g_r}, 
\eea
with 
\begin{equation}
 \frac{p(t)}{p_{ss}} = 1 - e^{- gt}(1 - gt)  ~~\textrm{if}~~  g_r = g_p = g  . \nonumber \\
\end{equation}

This expression can be simplified in the case of short-living mRNAs to $p(t)/p_{ss} \simeq (1-e^{-g_p t})$ as reported in~\cite{Rosenfeld02SI}.\newline 
The response time $T_{ON}$ is then defined by the equation $p(T_{ON})/p_{ss} = 0.5$. As can be seen in equations~\ref{on}, $T_{ON}$ does not depend on production rates ($k_r$ and $k_p$) but only on the half-lives of mRNAs and proteins. Since $T_{ON}$ for a sTF is independent of the final steady-state value of $p$,  if molecule half-lives are kept constant, it can be used as null model for comparison with iMSLs and tSLs at different levels of repression,  without the need of constraints on parameters.\newline
Analogously, the response time $T_{OFF}$ to a switch-off stimulus can be derived. In this case, the initial condition is the steady state given by a fully activated promoter $p(0)=p_{ss}$, and $k_r$ is set to zero at $t=0$. Again the dynamics depends only on the half-lives of 
mRNAs and proteins:

\bea
\frac{r(t)}{r_{ss}}& =& e^{-g_r t} \, , \nonumber \\
\frac{p(t)}{p_{ss}} &=& \frac{g_p e^{-g_r t} - g_r e^{-g_p t}}{g_p - g_r} \, , 
\eea

where

\begin{equation}
\frac{p(t)}{p_{ss}} = e^{- gt}(1 + gt) ~~\textrm{ if}~~ g_r = g_p = g. \nonumber \\
\end{equation}

The response time $T_{OFF}$ is given by the condition $p(T_{OFF})/p_{ss}=0.5$.

\subsection{Intronic miRNA-mediated self loop (iMSL)}

The deterministic description of the iMSL is given by equations 3 of the main text.
With the condition $q/h_{r}\gg 1$, required to have the target promoter
exposed to full activation, the transcription rate is at its maximum value $k_r$ and the steady-state solution can be easily found:

\bea
\label{steady-state-imsl}
s_{ss} &=& \frac{k_r}{g_s}  \nonumber \\
r_{ss} &=& \frac{k_r}{g_r}  \nonumber \\
p_{ss} &= &\frac{r_{ss}}{g_p} \frac{k_p}{1+s_{ss}/h}.
\eea

On the other hand, the dynamics and thus the response times can be extracted with numerical integration. 
These response times have been normalized in the main text, in particular in Figure 2, with the response times of a sTF (calculated as explained in the previous section), 
so as to evaluate the differences in the dynamics with respect to a constitutive transcription unit. 

\subsection{Transcriptional self regulation (tSL) }

The tSL dynamics is described by the two equations:

\bea
\label{det_prot}
\frac{dr}{dt} &=& k_r(q,p) -  g_r r  \nonumber \\
\frac{dp}{dt} &= & k_p r - g_p r,
\eea

where the transcription rate $k_{r}(q,p)$ is a product of two Michaelis-Menten-like functions: one corresponding to activation by  the TF $q$ and the other one taking into account the 
transcriptional self-repression. 
The choice of a simple product of functions implies the assumption of independent binding of the two regulators~\cite{Bintu05SI}, which is probably the most common situation.   
Therefore, the form of the transcription rate is:

\be 
k_r(q,p)  = k_r \left[\frac { q} {h_{r} +q}~ \frac{1} {1 + (\frac{p}{h_p})}\right].
\label{tsl-rate}
\ee

The condition $q/h_{r}\gg 1$ leads to the simplification $k_r (q,p) \simeq k_r(p) =k_r \frac{1} {1 + (\frac{p}{h_p})}$. 
In this case the  steady-state solution is:

\bea
\label{steady-state-stl}
& & r_{ss} = \frac{-g_p g_r h_p + \sqrt{g_p g_r h_p} \sqrt{g_p g_r h_p + 4 k_p k_r}}{2 g_r k_p}  \nonumber \\
& & p_{ss} = \frac{\frac{-g_p h_p + \sqrt{g_p h_p} \sqrt{g_p g_r h_p + 4 k_p k_r}}{\sqrt{g_r}}}{2 g_p} .
\eea

In order to compare the dynamics of tSLs with the iMSL one in un unbiased way, we impose a constraint on parameters in order to have the same steady state $p_{ss}$ for the target protein level. This can be simply done by equating
the values of $p_{ss}$ in  equations \ref{steady-state-stl} and \ref{steady-state-imsl} so as to extract the constraint on the parameter $h_p$ which  sets the repression strength in the tSL depending on the repression strength $h$ in the imLS:

\be
\label{constraint}
h_p = \frac{g_{s}^{2} h^{2} k_p}{g_p g_r (g_s h + k_r)}. 
\ee

Using this constraint, the response times for the tSL circuit can be evaluated numerically and directly compared with the response times of the iMSL circuit.

\section{Conditions for adaptation and Weber's law implementation}
\label{adaptation-sec}

As discussed in the main text, a necessary conditions to have perfect adaptation is the maintenance of a steady state independent of the input level, if this input is constant. In this way, the system can have a dynamical response to input changes while returning to its initial condition once the input level is steady for a sufficiently long time.\newline
In the case of iMSLs, a strong miRNA repression can make the circuit fulfill this condition. In the regime of strong repression $s/h \gg 1$, the translation rate simplifies to $k_{p}(s) \simeq h k_{p}/s$. The substitution of this expression in the equations describing the circuit dynamics (equations 3 of the main text), leads to a steady-state solution of the form: 

\bea
\label{adaptation_ss}
 s_{ss} &=& \frac{k_{r}(q)}{g_s}  \nonumber \\
 r_{ss} &=& \frac{k_{r}(q)}{g_r} \nonumber \\
 p_{ss} &=& \frac{k_p g_s}{ h g_r g_p}.
\eea

The steady state of the host-gene protein does not depend on the input level $q$, thus the iMSL circuit can in principle implement perfect adaptation.
\newline  

Weber's law requires additionally that the peak of the dynamical response depends only on the fold-change of the input. 
Introducing the further assumption of an approximately linear transcriptional activation (i.e. the amount of TFs $q$ is far from saturating the
target promoter), the transcription rate becomes $ k_{r}(q) \simeq  \frac{k_{r}}{h_r} q$ .   
Therefore, the two conditions of strong repression and approximately linear activation simplify the equations of the iMSL dynamics (equations 3 of the main text) to:

\bea
\label{FC}
\frac{ds}{dt} &=& \frac{k_r}{h_r} q -  g_s s\nonumber \\
\frac{dr}{dt} &=& \frac{k_r}{h_r} q -  g_r r\nonumber \\
\frac{dp}{dt} &=& h k_p \frac{r}{s} -  g_p p.
\eea

Assuming that the mRNA half-life is shorter than the other time scales in the system, a quasi-steady-state approximation can be used to further reduce the kinetic equations to:

\bea
\label{red}
\frac{ds}{dt} &=& \frac{k_r}{h_r} q -  g_s s \nonumber \\
\frac{dp}{dt} &=& \frac{k_p h k_r}{h_r g_r} \frac{q}{s} - g_p p  .
\eea 

We can now define the following dimensionless variables:

\bea
\label{adim}
& & t' = g_s t  \\
& & s' = \frac{g_s h_r}{k_r q_0} s  \nonumber \\
& & p' = \frac{g_r g_p}{g_s k_p h}  \nonumber \\
& & F = q / q_0  \nonumber \\
& & \phi = g_s / g_p ,
\eea
where the input stimulus is represented by a change in the TF concentration from a basal level $q_0$ to a new level $q$ ($F$ is the fold-change). Equations~\ref{red} can be thus rewritten as :

\bea
\label{FC2}
\frac{ds'}{dt'} &=& F - s' \nonumber \\
\phi \frac{dp'}{dt'} &=& \frac{F}{s'} - p' .
\eea

This reformulation shows that the dynamics of the target protein depends only on the fold-change $F$ in the input stimulus and not on its absolute value.  
Equations \ref{FC2} are the analogous of the equations presented in~\cite{Goentoro09SI} for the feed-forward loop circuit, adapted here to the iMSL case. Therefore, if the three conditions of strong repression, almost linear transcriptional activation and short mRNA half-life are satisfied, iMLs can implement Weber's law.

\section{Noise buffering: master equation and generating function approach}

\subsection{Intronic miRNA-mediated self loop (iMSL)}

This section  briefly describes the procedure to calculate analitically the coefficient of variation $CV_{x_i}$ for the molecular species $x_i$ 
involved in the  intronic miRNA-mediated self loop. The procedure can be similarly applied to the other two circuits (sTF and tSL).\\
For the stochastic analysis, the dynamics of transcription, translation and degradation of the TF is included explicitly. Therefore, the additional  variable $w$, 
representing the TF mRNA number is added in the system description, as well as its transcription rate $k_w$, translation rate $k_q$ and the degradation rates $g_w$ and $g_q$. 
In this way, the noise level 
in the input signal can be naturally modulated changing the relative contribution of transcription and translation to the $\langle q \rangle$ steady-state level, 
as described in details in~\cite{Osella11SI}.

The following master equation describes the evolution of the probability to find in a cell exactly ($w,q,s,r,p$) molecules at a given time $t$:

\bea
\label{master}
& & \partial_t P_{w,q,s,r,p} = ~~k_{w} (P_{w-1,q,s,r,p} - P_{w,q,s,r,p}) + k_q w (P_{w,q-1,s,r,p}-P_{w,q,s,r,p}) \nn \\
& & + k_r(q) (P_{w,q,s-1,r-1,p} - P_{w,q,s,r,p}) + k_p(s) r (P_{w,q,s,r,p-1}-P_{w,q,s,r,p}) \nn \\
& & + g_w \left[ (w+1)P_{w+1,q,s,r,p} - w P_{w,q,s,r,p} \right] + g_q \left[ (q+1)P_{w,q+1,s,r,p} - q P_{w,q,s,r,p} \right] \nn \\
& & + g_r \left[ (r+1)P_{w,q,s,r+1,p} - r P_{w,q,s,r,p} \right] + g_s \left[ (s+1)P_{w,q,s+1,r,p} - s P_{w,q,s,r,p} \right] \nn \\
& & + g_p \left[ (p+1)P_{w,q,s,r,p+1} - p P_{w,q,s,r,p} \right]  ,
\eea

where $k_r(q)$ and $k_p(s)$ have the functional form described in the main text. \newline
In order to evaluate the noise level at steady state, given by the coefficient of variation $CV_{x_i} \equiv \sigma_{x_i}/\langle x_i \rangle$, for each molecular species $x_i$,  it is necessary to find a closed expression for the 
first two moments of the above probability distribution at equilibrium.
To this aim, it is sufficient to linearize the regulation functions $k_r(q)$ and $k_p(s)$~\cite{Osella11SI,Komorowski09SI,Thattai01SI}, and apply the moment generating function approach to the resulting master equation at equilibrium~\cite{vanKampen81SI}. 
Even after the linearization, the system preserves a nonlinearity due to the term encoding the target translation, which still depends on both the number of mRNAs and miRNAs, but nonetheless the first two moments for the $p$ distribution can be calculated.
Using the linearization of the regulation functions:

\bea
k_{r} (q) &\simeq& k_{r} (q)|_{\langle q\rangle} + \partial_{q} k_{r} (q)|_{\langle q\rangle } (q - \langle q \rangle)  \nonumber \\
k_{p} (s) &\simeq& k_{p} (s)|_{\langle s \rangle} + \partial_{s} k_{p} (s)|_{\langle s\rangle} (s - \langle s \rangle) ,
\eea

the two rates can be redefined as:
\bea
k_{r} (q) &\simeq& k_{r}^{0} + q k_{r}^{1}  \nonumber \\
k_{p} (s) &\simeq& k_{p}^{0} - s k_{p}^{1}, 
\eea

with:

\bea
k^{r}_{0} &=& k_{r} (q)|_{\langle q\rangle} - \partial_{q} k_{r} (q)|_{\langle q\rangle} \langle q\rangle  \nonumber \\
k^{r}_{1} &=& \partial_{q} k_{r} (q)|_{\langle q\rangle}  \nonumber \\
k^{p}_{0} &=& k_{p} (s)|_{\langle s \rangle} + \partial_{s} k_{p} (s)|_{\langle s\rangle} \langle s\rangle  , \nonumber \\
k^{p}_{1} &=& - \partial_{s} k_{p} (s)|_{\langle s\rangle} ,
\eea

By defining the generating function: 
\be
F(z_1,z_2,z_3,z_4,z_5) = \sum_{w,q,s,r,p} z_{1}^w ~z_{2}^q ~ z_{3}^s~ z_{4}^r ~z_{5}^p ~P_{w,q,s,r,p} ,
\ee

and using the linearized regulation functions, equation \ref{master} can be converted in the following second-order partial differential equation:
\bea
\partial_{t} F = k_w (z_1 F - F) + k_q z_1 (z_2 \partial_{z_1} F - \partial_{z_1} F) + k_{r}^{0} (z_4 F - F)  \nonumber \\ 
+ k_{r}^{1} z_2 (z_4 \partial_{z_2} F - \partial_{z_2} F) + k_{p}^{0} z_4 (z_5 \partial_{z_4} F - \partial_{z_4} F)  \nonumber \\
- k_{p}^{1} z_3 z_4 (z_5 \partial_{z_{3},z_{4}} F - \partial_{z_{3},z_{4}} F) + g_w (\partial_{z_1} F - z_1 \partial_{z_1} F)  \nonumber \\
+ g_q (\partial_{z_2} F - z_2 \partial_{z_2} F) + g_s (\partial_{z_3} F - z_3 \partial_{z_3} F)   \nonumber \\
+ g_r (\partial_{z_4} F - z_4 \partial_{z_4} F) + g_p (\partial_{z_5} F - z_5 \partial_{z_5} F) .
\label{partial}
\eea
                
The differentiation of \ref{partial} at the steady state 
leads to equations for successively higher moments thanks to the following properties of the moment generating function: 
$F|_1=1$, ${\partial_{z_i}} F = \langle x_i\rangle$ and 
${\partial^{2}_{z_i}} F = \langle x_i^2\rangle - \langle x_i\rangle$ (where $|_1$ denotes the evaluation of $F$ at $x_i = 1$ for all $i$). 
Differentiating up to the fourth moments, the analytical expression for $CV_{x_i}=\frac{\sigma_{x_i}}{<x_i>}$ can be obtained 
(see~\cite{Osella11SI} for a more exhaustive and detailed analysis), and thus also the noise level of the host-gene protein product used in the main text.

\subsection{Simple transcriptional unit (sTF)}
The stochastic analysis of the sTF can be performed as explained in the previous section. 
We just report here the corresponding master equation: 
\bea
\frac{\partial P_{w,q,r,p}}{\partial t} = k_w (P_{w-1, q, r, p} - P_{w,q,r,p}) + k_q w (P_{w,q-1,r,p} - P_{w,q,r,p}) \nonumber \\
+ k_r (q) (P_{w,q,r-1,p} - P_{w,q,r,p}) + k_p r (P_{w,q,r,p-1} - P_{w,q,r,p}) \nonumber \\
+ g_w [(w+1) P_{w+1,q,r,p} - w P_{w,q,r,p}] + g_q [(q+1) P_{w,q+1,r,p} - q P_{w,q,r,p}] \nonumber \\
+ g_r [(r+1) P_{w,q,r+1,p} - r P_{w,q,r,p}] + g_p [(p+1) P_{w,q,r,p+1} - p P_{w,q,r,p}] .
\eea

\subsection{Transcriptional self-regulation (tSL)}

The master equation for the transcriptional self-regulation (tSL) is:
\bea
\frac{\partial P_{w,q,r,p}}{\partial t} = k_w (P_{w-1, q, r, p} - P_{w,q,r,p}) + k_q w (P_{w,q-1,r,p} - P_{w,q,r,p}) \nonumber \\
+ k_r (q,p) (P_{w,q,r-1,p} - P_{w,q,r,p}) + k_p r (P_{w,q,r,p-1} - P_{w,q,r,p}) \nonumber \\
+ g_w [(w+1) P_{w+1,q,r,p} - w P_{w,q,r,p}] + g_q [(q+1) P_{w,q+1,r,p} - q P_{w,q,r,p}] \nonumber \\
+ g_r [(r+1) P_{w,q,r+1,p} - r P_{w,q,r,p}] + g_p [(p+1) P_{w,q,r,p+1} - p P_{w,q,r,p}] .
\eea

In order to calculate $CV_p$ with the moment generating function approach, it is necessary to define the linearization of the function $k_r(q,p)$ shown  in equation~\ref{tsl-rate}.
As described in~\cite{Osella11SI}, we can linearize it as:

\bea
k_r(q,p) & \simeq & k_r(q,p)|_{\langle q\rangle,\langle p\rangle} + \partial_{q} k_r(q,p)|_{\langle q\rangle,\langle p\rangle} (q-\langle q\rangle) \nn\\
	&+ & \partial_{p} k_r(q,p)|_{\langle q\rangle,\langle p\rangle} (p-\langle p\rangle), 
\label{linearization-protein}
\eea

and thus obtain a transcription rate of the form $k_r(q,p)  \simeq k_{r}^0 + k_{r}^1 q -k_{r}^2 p$. 
Using this linearization and the moment generating function approach,  the analytical expressions of $\langle p\rangle$ and $CV_p$ can be obtained as described in previous sections.

\section{Relation with other modeling strategies of miRNA-mediated regulation}

\subsection{Molecular titration model}
A mathematical model of miRNA-mRNA interaction was previously proposed to describe sRNA regulation in bacteria~\cite{Levine07SI}.
This model takes into account the physical coupling between miRNAs and mRNAs explicitly with a simple titration mechanism: 
a miRNA can form a complex with a target mRNA, degrade it and then eventually be available again to target other mRNAs. A parameter $\alpha$ is introduced to measure 
the probability of miRNA recycling after target degradation induced by mRNA-miRNA coupling. Thus, $\alpha$ represents the degree of ``catalyticity'' of miRNA regulation, 
with $\alpha=0$ for perfect catalytic action, while $\alpha=1$ for stoichiometric action. \newline
Applying this modeling strategy to the iMSL circuit, the following system of differential equations is obtained:

 \bea
\frac{ds}{dt} &=& k_r(q) -  g_s s -(k_{+} r s - k_{-} c) + (1-\alpha) \beta c  \nonumber \\
\frac{dr}{dt} &=& k_r(q) - g_r r - (k_{+} r s - k_{-} c) \nonumber \\
\frac{dc}{dt} &=&  k_+ {r s} - k_{-} c -\beta c\nonumber\\
\frac{dp}{dt} &=& k_p r -  g_p p .
\label{general}
\eea

In this equations, $c$ is the concentration of the miRNA-mRNA complex,  $k_{+}$  is the probability of miRNA-mRNA
association and $k_{-}$ the probability of dissociation of the complex $c$, that can degrade with rate $\beta$.\newline
An analogous  model  of miRNA regulation have been used to describe the results of single-cell experiments in mammalian cells~\cite{Mukherji11SI},
 with the additional assumption of slow miRNA turnover, thus neglecting the dynamics of miRNA transcription and degradation.

\subsection{Relations between titration model and phenomenological models based on Michaelis-Menten functions}

If the coupling of miRNAs and mRNAs is fast, or if the interest is on steady state properties, the $c$ dynamics can be equilibrated in equations~\ref{general}:

\bea
\frac{ds}{dt} &=& k_r(q) -  g_s s -\alpha k_{rs} r s  \nonumber \\
\frac{dr}{dt} &=& k_r(q) - g_r r -  k_{rs} r s  \nonumber \\
\frac{dp}{dt} &=& k_p r -g_p p,
\label{reducedhwa}
\eea

 with $k_{rs}= \beta k_{+} /( k_{-} +\beta)$. miRNA regulation is often distinguished from sRNA one because of the efficient recycling (i.e. catalytic interaction). In particular, in the case of 
$\alpha \simeq 0$, the molecular concentrations at steady state are:

\bea
r_{ss} &=&\frac{k_r}{g_r} \frac{1}{1 + k_{rs} s/g_r}   \nonumber \\
p_{ss} &=& k_p  \frac{k_r}{g_r}  \frac{1}{1+k_{rs}/g_r ~s} = r_{0} k_p \frac{1}{1+k_{p} s/h},
\eea

where $r_{0}=k_r(q)/g_r$ is the steady state for a constitutive mRNA, while $h=g_r/k_{rs}$. This expression for the target protein concentration
 shows the equivalency to the model of miRNA inhibition of target translation used in the main text. In fact, the same steady state can be obtained using an effective Michaelis-Menten function of miRNA concentration as target translation rate (as in equations 3 of the main text),  with an effective dissociation constant $h=g_r/k_{rs}$. 
Therefore, in the limit of high miRNA recycling the steady state properties of a titration model are completely equivalent to an effective model of nonlinear miRNA action on target translation rate.\newline
\newline
The miRNA regulation can also be modeled using an effective nonlinear function in the mRNA degradation term, thus assuming that miRNA regulation acts mainly on the stability of 
mRNAs rather than on their translation efficiency. In this case, the equations describing the iMSL circuit dynamics are:

\bea
\frac{ds}{dt} &=& k_r(q) -  g_s s   \nonumber \\
\frac{dr}{dt} &=& k_r(q) - (g_r + g_{max}  \frac{s}{h+s}) r \  \nonumber \\
\frac{dp}{dt} &=& k_p r -g_p p.
\label{degradation}
\eea

The miRNA action is represented by adding to the basal rate of mRNA degradation $g_r$ (in absence of miRNAs) an increasing  function of miRNA concentration, where $g_{max}$ is the maximum possible increase of the degradation rate (if $ s \rightarrow \infty, g_{r} (s) \rightarrow g_r + g_{max} $) and $h$ is the dissociation constant of miRNA-mRNA interaction. 
It's easy to see that in the case of strong enough repression, for which $s/h \gg 1$, the equation for $r$ in ~\ref{degradation} can be recasted as:

\begin{equation}
\frac{dr}{dt} = k_r(q) - g_r  r  - \frac{g_{max}} {h} r s,
\end{equation}

 making clear the relation between this description and the titration model with fast binding/unbinding of mRNA and miRNAs and high cataliticity, 
i.e. equations~\ref{reducedhwa} with $\alpha \simeq 0$ and $k_{rs}=g_{max}/h$.

\subsubsection{Adaptation and Weber's law conditions in the titration model}

As shown in section~\ref{adaptation-sec}, the iMSL can perform adaptation in the regime of strong miRNA-mediated repression. In the context of the titration model with high catalyticity ($\alpha \simeq 0$), strong repression implies that the degradation of mRNAs is dominated by miRNA regulation. Therefore, we can approximate equations~\ref{reducedhwa} with:

\bea
\frac{ds}{dt} &=& k_r(q) -  g_s s  \nonumber \\
\frac{dr}{dt} &\simeq& k_r(q) -  k_{rs} r s  \nonumber \\
\frac{dp}{dt} &=& k_p r -g_p p.
\label{reducedhwarep}
\eea

The steady state solution for the target protein is:

\be
p_{ss} = \frac{k_p g_s}{k_{rs} g_p}.
\ee

This expression is independent on the input level $q$, showing that the condition for adaptation is satisfied in the strong repression regime also for this alternative modeling strategy of miRNA regulation, if the miRNA recycling is sufficiently efficient as it is expected for miRNA regulation in higher eukaryotes. 
Given the steady-state equivalency (shown in the previous section) between the titration model with $\alpha \simeq 0$ and a phenomenological model of mRNA degradation induction, the addition of iMSLs to the list of adaptive circuits seems robust and model-independent.\newline 
Starting from equations \ref{reducedhwarep}, in which there is also the implicit assumption of fast mRNA-miRNA binding/unbinding, also the conditions for Weber's law implementation can be examined in the context of the titration model. As previously discussed, the additional requirements with respect to adaptation are an almost linear transcriptional activation and a fast mRNA dynamics. With these two constraints the dynamic equations become:

\bea
\frac{ds}{dt} &=& \frac{k_r}{h_r} q  -  g_s s  \nonumber \\
\frac{dp}{dt} &=& \frac{k_p k_r}{h_r k_{rs}} \frac{q}{s} -g_p p.
\eea

These two equations have exactly the same form of equations~\ref{red}, thus can be similarly reformulated in terms of adimensional variables, showing that the $p$ dynamics depends only on the input fold-change. 

\subsubsection{Comparison of the response times for different models of miRNA-mRNA interaction.}

A direct comparison of the response times for the different models of miRNA-mRNA interaction is made difficult by the higher number of parameters that are present in the titration model (equations~\ref{general}) with respect to the phenomenological model presented in the main text. The way in which the two models can be constrained to have the same level of target protein at equilibrium, in order to make an unbiased comparison, is indeed quite arbitrary. In particular,  the timescale of the binding/unbinding of mRNAs and miRNAs in the $c$ complex can strongly influence the dynamics.\\
For example,  for fast binding/unbinding (as in equation~\ref{reducedhwa} and equivalently in equations~\ref{degradation}), the iMSL reduces the time required to switch-on the host-gene expression, but also accelerates the switch-off dynamics. The reduced effective mRNA half-life drives a fast drop in mRNA concentration, and thus of proteins. Therefore, in this conditions the iMSL is not effective in keeping the ON-state robust with respect to fluctuations in the activator level. However, the opposite case of a long-living mRNA-miRNA complex have dynamical properties more similar to those of the phenomenological model presented in the main text. The simplified situation of mRNA sequestration in long-living miRNA-mRNA complexes from which mRNAs cannot be translated, but mRNAs and miRNAs degrade with their natural rates, can be considered for a comparison. The iMSL dynamics in this case is described by the equations:

\bea
\frac{ds}{dt} &=& k_r(q) -  g_s s -(k_{+} r s - g_{r} c)  \nonumber \\
\frac{dr}{dt} &=& k_r(q) - g_r r - (k_{+} r s - g_{s} c) \nonumber \\
\frac{dc}{dt} &=&  k_+ {r s} - (g_r + g_s) c \nonumber\\
\frac{dp}{dt} &=& k_p r -  g_p p .
\label{sequester}
\eea

This model can be easily constrained to have the same $p_{ss}$ solution of the model presented in the main text, acting on the parameter $k_{+}$. 
The results of the analysis of the response times are qualitatively equivalent to those presented in the main text, although a miRNA-mediated repression of translation seems quantitatively more efficient in locking the ON-state of the host-gene expression. \newline 
Therefore, while the acceleration of the host-gene activation is a result independent of the type of miRNA repression, the delayed switch-off kinetics is expected to be observed for miRNAs repressing target translation or miRNAs that can bind mRNAs in sufficienlty stable complexes. A stoichiometric repression based on coupled miRNA-mRNA degradation, 
like the one reported for sRNAs in bacteris~\cite{Levine07SI}, or a nonlinear induction of mRNA degradation could instead change the switch-off dynamics. A better experimental understanding of the mechanisms of miRNA-mRNA interaction in the specific case in analysis and  measurments of the model parameters are thus required to fully address the details of the dynamics of miRNA-mediated circuits with a  quantitative mathematical description.

\section{Bioinformatic analysis - Supplementary tables}

\begin{figure}[h]
\includegraphics[width=1\textwidth]{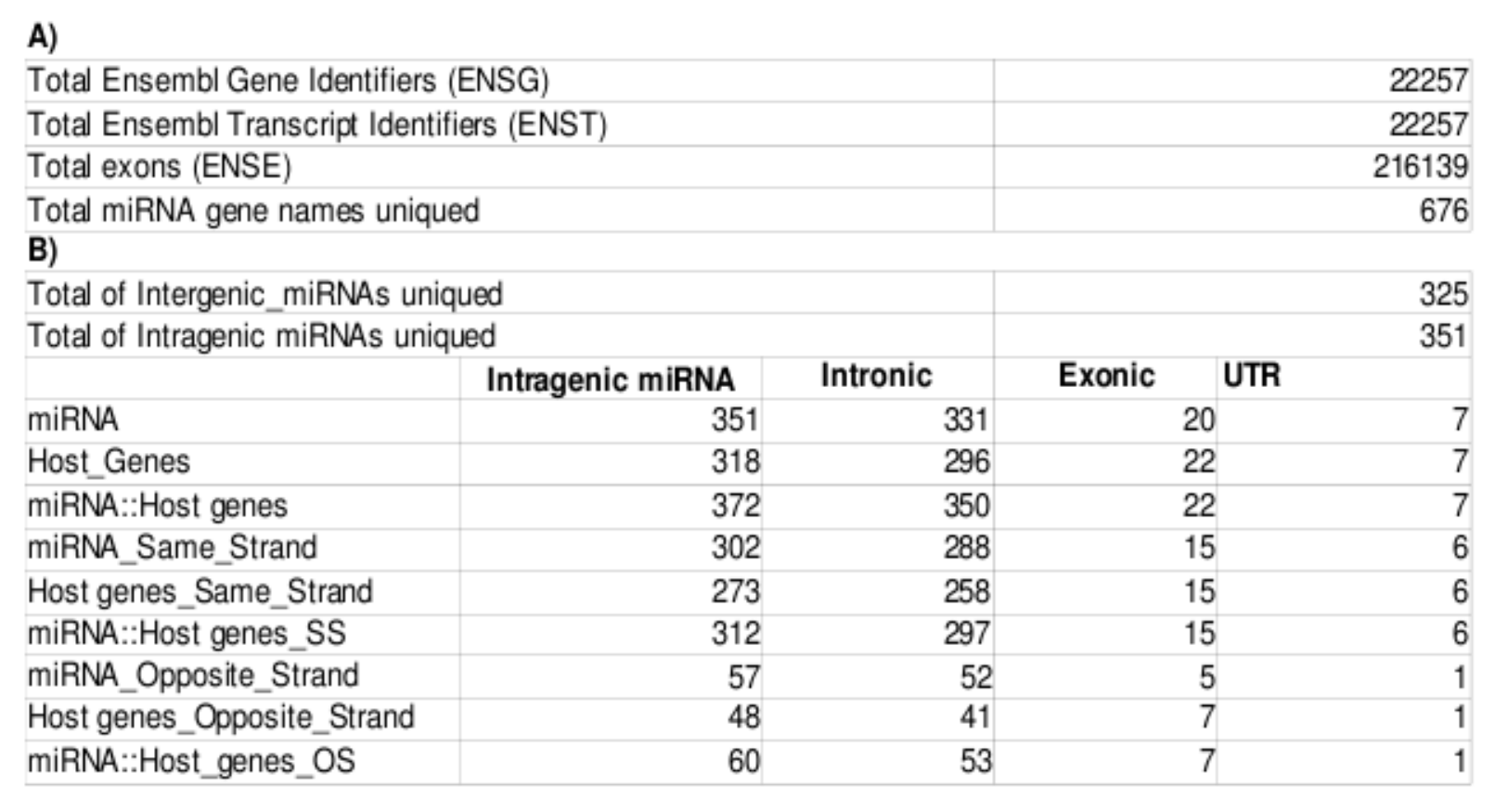}
\caption{
Table 1S. (A) Description of human known protein coding genes and human miRNA datasets. For each ENSG, we considered the longest Ensembl transcript ID (ENST). Data from Ensembl v.57 that include miRBase v.13. 
(B) Classification of human miRNAs with respect to their host genes. Depending on the genomic location of human miRNAs, we considered as intergenic miRNAs whose genomic position were found distant from annotated genes, while intragenic miRNAs 
whose located within a transcript (annotated as ``host gene”). Afterward, intragenic miRNAs were further subdivided into intronic and exonic. 
An intragenic miRNA was called exonic if its genomic coordinates overlap with genomic coordinates of any exon in the database, and was labeled intronic otherwise. In addition, intragenic miRNAs can be classified depending on whether they are on 
the same strand (SS) or on the opposite strand (OS) of their host gene. All UTR miRNAs were found to overlap the exon regions.
}
\label{table1}
\end{figure}


\begin{sidewaysfigure}[h]
\begin{centering}
\includegraphics[scale=0.8]{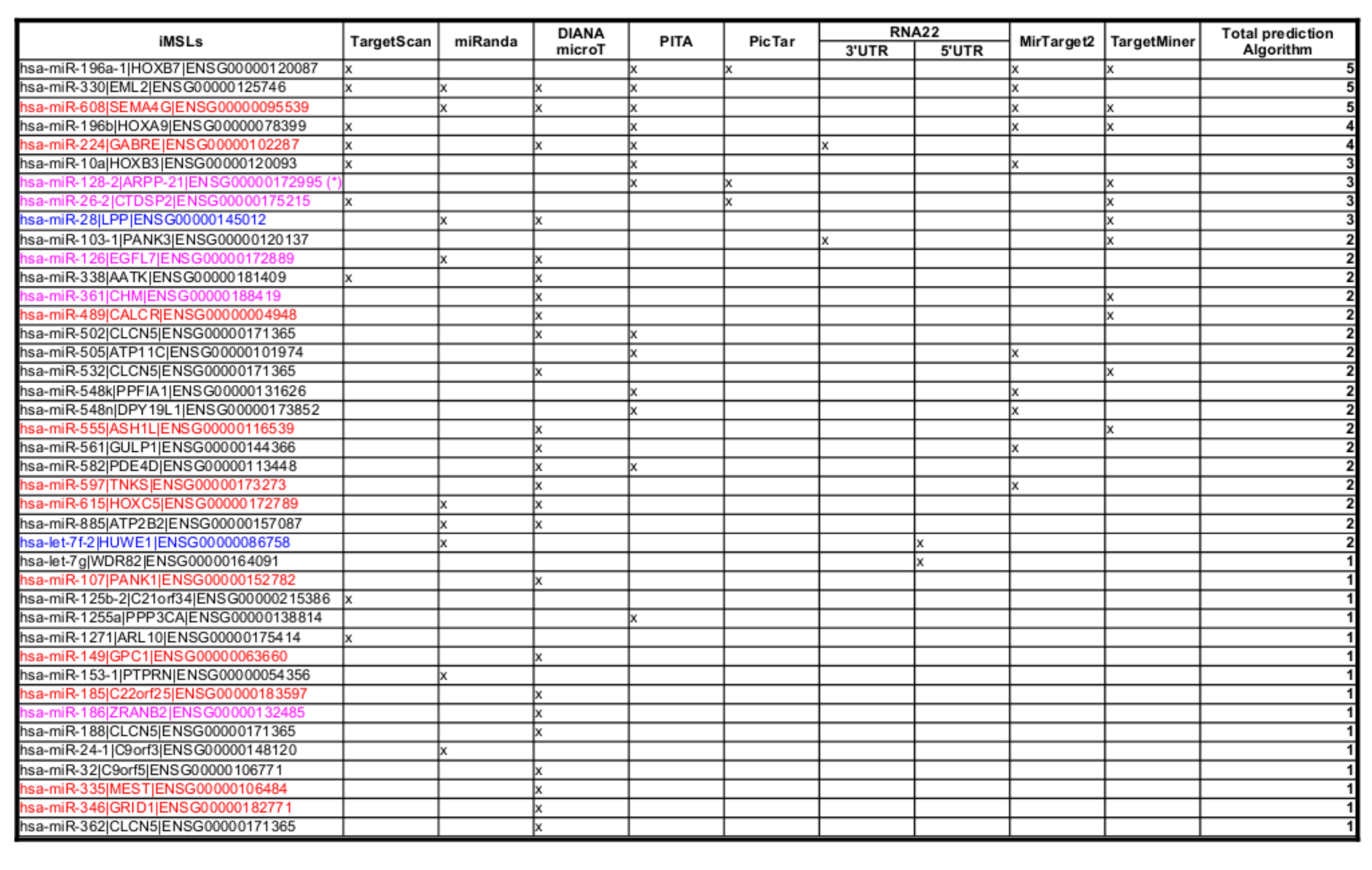}
\caption{Table 2S. List of intronic miRNA-mediated self loops (iMSLs). For each iMSL we highlighted with ``x" the correspondent dataset where they were found, with blue color if they were reported by~\cite{Tsang07SI}, 
with red if they were found by~\cite{Megraw09SI} and with fuchsia if they were reported in both these two studies. The only iMSL experimentally validated was highlighted with ($*$) symbol.}
\label{table2a}
\end{centering}
\end{sidewaysfigure}


\begin{sidewaysfigure}[h]
\begin{centering}
\includegraphics[scale=0.8]{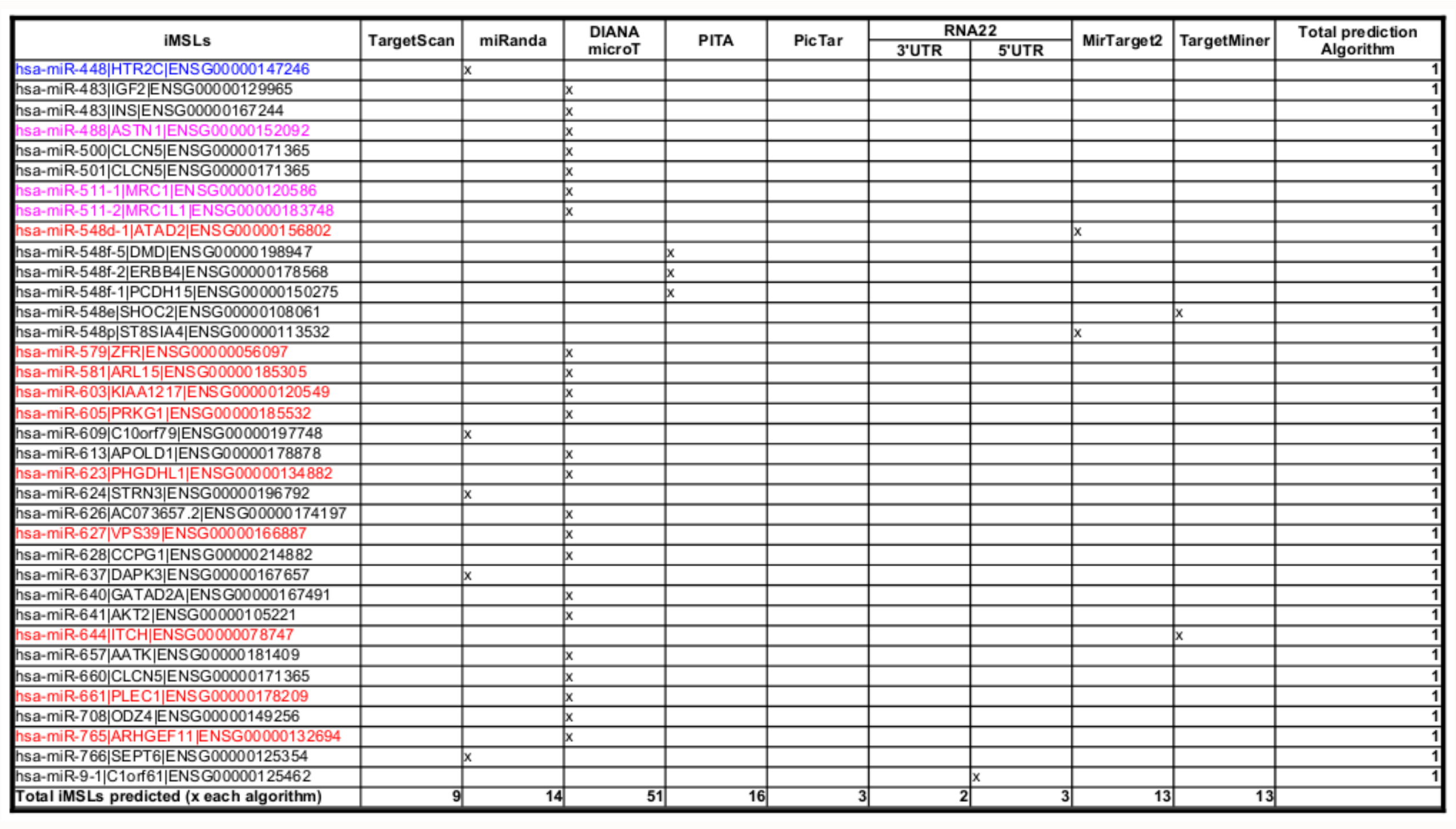}
\caption{
Sequel of Table 2S.
}
\label{table2b}
\end{centering}
\end{sidewaysfigure}

\end{bmcformat}
\end{document}